\documentclass{aa}

\usepackage{graphicx}
\usepackage{txfonts}

\newcommand{\msun}{\mbox{M$_\odot$}}
\newcommand{\kms}{\mbox{km s$^{-1}$}}
\newcommand{\dg}{\mbox{$^\circ$}}
\newcommand{\am}{\mbox{$^{\prime}$}}
\newcommand{\as}{\mbox{$^{\prime\prime}$}}

\newcommand{\nhp}{\mbox N$_{2}$H$^{+}$}

\defcitealias{Dun15}{D15}
\begin{document}

   \title{Distribution of Serpens South protostars revealed with ALMA}

\author{Adele L. Plunkett\inst{1}
  \and Manuel Fern\'{a}ndez-L\'{o}pez\inst{2} 
     \and H\'{e}ctor G. Arce \inst{3}
     \and Gemma Busquet\inst{4}
     \and Diego Mardones\inst {5}
     \and Michael M. Dunham \inst{6,7}} 

\offprints{A. L. Plunkett, \email{aplunket@eso.org}}

\institute{European Southern Observatory, Av. Alonso de C\'{o}rdova 3107, Vitacura, Santiago, Chile 
  \and Instituto Argentino de Radioastronom\'{i}a, CCT-La Plata (CONICET), C.C.5, 1894, Villa Elisa, Argentina 
    \and Department of Astronomy, Yale University, P.O. Box 208101, New Haven CT 06520, USA
    \and Institut de Ci\`encies de l'Espai (IEEC-CSIC), Campus UAB, Carrer de Can 
   Magrans, S/N E-08193, Cerdanyola del Vall{\`e}s, Catalunya, Spain
    \and Departamento de Astronom\'{i}a, Universidad de Chile, Casilla 36-D, Santiago, Chile 
    \and Department of Physics, State University of New York at Fredonia, Fredonia, NY 14063, USA
        \and Harvard-Smithsonian Center for Astrophysics, 60 Garden Street, MS 78, Cambridge, MA 02138, USA 
} 

   \date{Accepted to A\&A March 2018}

  \abstract{\textbf{Context. }Clusters are common sites of star formation, whose members display varying degrees of mass segregation.  The cause may be primordial or dynamical, or a combination both.  If mass segregation were to be observed in a very young protostellar cluster, then the primordial case can be assumed more likely for that region.
 
  \textbf{Aims. } We investigated the masses and spatial distributions of pre-stellar and protostellar candidates in the young, low-mass star forming region Serpens South, where active star formation is known to occur along a predominant filamentary structure. Previous observations used to study these distributions have been limited by two important observational factors: (1) sensitivity limits that leave the lowest-mass sources undetected, or (2) resolution limits that cannot distinguish binaries and/or cluster members in close proximity.
 
  \textbf{Methods. } Recent millimeter-wavelength interferometry observations can now uncover faint and/or compact sources in order to study a more complete population of protostars, especially in nearby ($D<500$\ pc) clusters. Here we present ALMA observations of 1 mm (Band 6) continuum in a $3 \times 2$\ arcminutes region at the center of Serpens South. Our angular resolution of $\sim1\as$\ is equivalent to $\sim400$\ au, corresponding to scales of envelopes and/or disks of protostellar sources.
 
 \textbf{Results. }  We detect 52 sources with 1 mm continuum, and we measure masses of $0.002 - 0.9$\ solar masses corresponding to gas and dust in the disk and/or envelope of the protostellar system.  For the deeply embedded (youngest) sources with no IR counterparts, we find evidence of mass segregation and clustering according to: the Minimum Spanning Tree method, distribution of projected separations between unique sources, and concentration of higher-mass sources near to the dense gas at the cluster center.
 
 \textbf{Conclusions. }  The mass segregation of the mm sources is likely primordial rather than dynamical given the young age of this cluster, compared with segregation time.  This is the first case to show this for mm sources in a low-mass protostellar cluster environment.}

\keywords{Stars: formation -- Stars: protostars -- Submillimeter: stars -- Methods: observational -- Techniques: interferometric  }

   \maketitle

\section{Introduction}

Stars in the Milky Way preferentially form in clustered environments \citep{Lad03,Bre10}.  Moreover, in many young stellar clusters containing high-mass stars, the most massive stars concentrate toward the center of the cluster according to an effect called ``mass segregation'' \citep[e.g.][]{Zin93}.  The physical origin of this effect is uncertain due to intrinsic observational challenges (for example, mass segregation is a relatively short process, and protostars are deeply embedded in early stages), as well as a lack of clear understanding of the structure and dynamics of the large-scale motions governing cloud dynamics. In essence, there are two possibilities to explain mass segregation: dynamical or primordial \citep[or some combination of both, such as primordial segregation followed by a dynamical post-segregation][]{2009MNRAS.396.1864M}.  Moreover, it is unclear whether this process should apply to low(er)-mass stellar clusters.

Simulations have shown that, after some time, two- and/or multi-body dynamical interactions between stars can lead to partial mass segregation down to a ``limiting mass'' \citep[e.g.][]{All09a,Pan13}, with the more-massive stars becoming mass segregated earlier than less-massive stars. An alternative scenario suggests a primordial origin for mass segregation \citep[e.g.][]{Bon98}. In this case, the position of the stars in the cluster mainly reflects the gas and dust distribution of the natal cloud, which formed the most massive stars near the center of the gravitational potential; simulations were presented by e.g.~\citet{Bon98,Bon06}, while near-IR observational evidence have been presented by e.g.~\citet{Tes97,Tes98b,Hil98,Mas00,Mas03}. In summary, a cluster will always be mass segregated down to a limiting mass after some time (after a complete relaxation time it will become mass segregated down to the average mass of its stellar members), but before this occurs, a cluster can only be mass segregated if primordial segregation is in effect.

Observational studies have shown evidence for mass segregation in the young stellar groups in Taurus, Lupus3, ChaI, and IC348 \citep{2011ApJ...727...64K}, although another method found inverse segregation instead in Taurus \citep{2011MNRAS.412.2489P}.  In the sample of IC348, Serpens, NGC1333, and L1688, \citet{2008MNRAS.389.1209S} found signs of mass segregation in the older clusters (the former two) but not in the younger clusters (the latter two), suggesting dynamical effects.  Additional mass segregation studies have targeted young, massive Galactic stellar clusters and associations including ONC \citep{Hil98,All09b}, NGC3603 \citep{Pan13}, Westerlund 1 \citep{Lim13}, Trumpler 14 \citep{San10}, and Berkeley 94 and Berkeley 96 \citep[to different degrees of segregation][]{2013MNRAS.435..429D}, to mention a few.  No evidence of mass segregation in the dynamically young clusters Rho Ophiuchus \citep{2012MNRAS.426.3079P} nor Cyg 0B2 \citep{2014MNRAS.438..639W} were found.  These studies were targeting young \textit{stars} using primarily optical/IR and/or X-ray selection techniques, and therefore they do not include the most embedded, youngest members.  Simulations also show varying degrees (or lack thereof) of primordial mass segregation \citep[e.g.][]{All10,2011MNRAS.416..541M,Gav17,2013MNRAS.432..986P}. Young massive \textit{proto}stellar clusters have been studied (at millimeter wavelengths) in a couple of cases \citep[see][]{Che17,Cyg17}, but a similar analysis is still lacking for more regions where pre-stellar or protostellar sources in the earliest stages are surrounded by disks and/or envelopes, and especially in less massive environments.  The low-mass, protostellar case is the focus of our study.

\subsection{ Serpens South}

\object{Serpens South Cluster} was discovered by inspecting \textit{Spitzer}\ observations in the Gould Belt survey \citep{Gut08}.  It was shown to have a high ratio of protostars to young stellar objects (YSOs), known as the protostar fraction, which is indicative that the cluster is very young with recent active star formation.  Based on a carbon-chain molecule study, \citet{Fri13} suggested an age of $2\pm1\times10^5$\ yr.  Due to its young age, this is a feasible region to test whether the distribution of protostars is primordial, or whether enough time has passed for significant dynamical evolution or disruption.

Several studies have presented observations of star formation tracers in Serpens South.  These include dust continuum maps \citep{Mau11,Kon15} and molecular line emission maps \citep{Kir13,Tan13,Fer14,Plu15a}.  Catalogs of YSOs have been presented by \citet{Dun15,Get17} using mid-IR and X-ray point sources, and for these sources evolutionary classifications are possible according to spectral energy distribution (SED) fitting of IR bands.  We use especially the catalog of \citet[][hereafter D15]{Dun15} for classifying sources in this work.

We focus on the central cluster region, where the protostar fraction reaches 91\% (throughout we consider Class 0/I and flat-spectrum sources as protostars).  Within a $\sim6$\ arcsec$^2$\ region with mass of 155.3\msun\ \citep[based on the N(H$_2$) map of ][]{Kon15} reside some $\sim$50 protostars.  We have previously mapped this region using millimeter-wavelength interferometry with CARMA with an angular resolution of $\sim5\arcsec$\ (equivalent to spatial resolution $\sim2200$\ au).  At the resolution and sensitivity of CARMA, we detected seven continuum sources, including several with irregular morphologies that hinted at binary or multiple systems \citep{Plu15a}.  A number of the sources are deeply embedded and only visible at millimeter wavelengths, and therefore they were not identified in IR source catalogs.  Hence, a full census of the protostars in this region requires sensitivity and resolution only recently accessible with ALMA.

Serpens South is one of only a few active star forming regions located at a distance less than $\sim500$\ pc. It is located 3 degrees south of Serpens Main, and it is embedded within the Aquila Rift along with the nearby region W40.  While no distance measurement has been made for a member of Serpens South, it has commonly been assumed that it is at the same distance as Serpens Main, and closer than W40.  For a more extensive discussion of distance, we refer the reader to \citet{Ort17}.  Briefly, since the discovery of Serpens South, different authors have assumed a range of distances, most commonly citing a relatively nearby distance of $260\pm37$\ pc \citep{Str96} that was determined using photometry of Serpens Main, and consistent with that of the larger sample from \citet{Str03}.  Later, Very Long Baseline Interferometry (VLBI) trigonometric parallax by \citet{Dzi10,Dzi11} provided a farther distance to Serpens Main of $415 - 419$\ pc.  Finally, multi-epoch Very Long Baseline Array (VLBA) observations by \citet{Ort17} provided the distance to Serpens/Aquila molecular complex of $436.0\pm9.2$\ pc, which we adopt for Serpens South in our study.

\section{Observations and data} 
\subsection{ALMA Observations}\label{sec:obs}
We observed the region Serpens South with ALMA (\#2012.1.00769.S). The observations where made with the 12m array using 32 antennas (baselines 15-437 m) and the 7m array using 9 antennas  (baselines 8-48 m) during 2014 January - June.  We made mosaic maps using 137 pointings and 53 pointings with the 12m array and 7m array, respectively.  Time on source during each execution block to make the maps were 41 minutes (times two executions) for the 12m array and 18-34 minutes (times 8 executions) for the 7m array.  

The observations were made in Band 6, and continuum maps were made by selecting line-free spectral channels in four basebands centered at 230.588 GHz, 231.45 GHz, 220.329 GHz, and 219.56 GHz.  These basebands had bandwidths of 234 MHz, 468 MHz, 234 MHz, and 234 MHz, respectively, and the total bandwidth of the continuum image (line-free channels) sums to 858 MHz.  The mean frequency of line-free channels is 225.758 GHz, which corresponds to a wavelength of 1.3 mm.  Molecular lines were included in the spectral setup, but the focus of this paper is the continuum map; further analysis of molecular line maps will be presented in the future.  One subset of data from the same project is included in \citet{Plu15b}.

Calibrators were dynamically chosen from standard catalogs.  Bandpass calibrators used were: J1733-1304, J1751+0939, J1550+0527.  Flux calibrators used were: Ceres, J1751+096, Mars, Neptune, Titan. J1751+0939 was used as phase calibrator.

We combined observations from the 12m and 7m arrays in the CLEAN process.  In the CASA tclean task, we used multi-frequency synthesis and a robust weighting parameter of 0.5 to make the continuum image.  We masked interactively, drawing CLEAN boxes around the emission and reviewing residuals before reiterating.  Finally we corrected for primary beam.  Our map has a synthesized beam HPBW of $0.96\arcsec\times 0.59\arcsec$, equivalent to a linear resolution of $419\times 256$\ au at a distance of 436 pc.  The maximum recoverable scale is $\sim29\arcsec$, or 0.06 pc (for the 7m array in Band 6).  The resulting mosaic map covers $3\arcmin$\ in declination and $2\arcmin$\ in right ascension (equivalent to $0.38\times 0.25$\ pc) within 30\% power of the antenna response; the map is shown in Figure \ref{fig:map}. \footnote{FITS data are available at the CDS via anonymous ftp to \url{cdsarc.u-strasbg.fr} (130.79.128.5) or via \url{http://cdsweb.u-strasbg.fr/cgi-bin/qcat?J/A+A/}.}

The rms noise level of the image is not constant within the 30\% antenna response of the mosaic due to dynamic range effects. In particular, in the central region where the brightest millimeter sources are found the rms noise level is about two times the rms in the outskirts of the image, varying between 0.3 and 0.8 mJy beam$^{-1}$.  The dynamic range (peak emission / minimum rms) considering the brightest millimeter source is $\sim 200-400$. 

\subsection{Supplementary data to study dense gas} \label{sec:extradata}

\nhp\ was observed with CARMA\footnote{Combined Array for Research in Millimeter-wave Astronomy} toward the Serpens South molecular cloud during the CLASSy project\footnote{CARMA Large Area Star Formation Survey \citep{Sto14}} campaign and the data were presented in \citet{Fer14}. The observations combine CARMA cross- and auto-correlations (i.e., single dish and interferometer data) giving a $\sim 7\arcsec$ resolution while including the zero-spacing information, thus recovering the extended spatial distribution of the gas emission. The correlator setting included a 8~MHz band centered at the rest frequency of the \nhp\ (J=1-0) line ($\nu_0=93.173704$~GHz) with 159 channels of 0.16 \kms. The \nhp\ is the preferred molecular tracer for the quiescent gas of the Interstellar Medium filaments. These data show the kinematics and spatial distribution of the well-known filamentary structure of the Serpens South cloud. At the intersection of the filament network lies a central hub, a strong \nhp\ emitter harboring the central cluster. Figure \ref{fig:map} presents the location of the Serpens mm cluster detected with ALMA relative to the CARMA \nhp\ (J=1-0) emission. We also use \nhp\ integrated intensity, velocity centroid and velocity dispersion images to extract the kinematical information of the line-of-sight gas at the location of the mm continuum sources (see Figure \ref{fig:massvdist} and \S \ref{sec:densegas}).  To obtain the velocity dispersion map \citep[see][for details on the procedure]{Sto14} the seven \nhp (1-0) hyperfine components were modeled simultaneously using Gaussians with the same dispersion and excitation conditions in pixels with enough signal-to-noise ratio.  

\section{Results}\label{sec:results}

\subsection{Millimeter continuum source identifications}
The mm-continuum emission detected with ALMA pertains to the dense dusty material associated with pre-stellar condensations or protostars in the process of collapse and disk-formation.  We show the ALMA mm-continuum map of the central cluster of Serpens South in Figure \ref{fig:map}.  As we explained in Section \ref{sec:obs}, rms noise level is not constant in the image, and in order to identify the millimeter sources we take this variation into account by changing the $5\sigma$ threshold detection limit accordingly throughout the image.

We identified 52 mm sources in our map (see Table \ref{tab:sources}); hereafter we use the term ``mm source'' to indicate an emission feature identified using the following method, and we label the sources as ``serps\#\#'', in order of increasing right ascension (note that in this list are also included IR sources from D15 with no mm continuum detection).  To identify mm sources, first we made two-dimensional (2D) Gaussian fits where emission exceeded $5\sigma$\ and where a local emission peak was seen by eye; fits were made using the task imfit in MIRIAD \citep{Sau95}.  In Figure \ref{fig:map_zoom} (especially panels a, c, d) one can appreciate the detail of several particularly crowded regions, yet generally a 2D Gaussian fit could be made around local emission peaks.  

At this stage, emission near two pairs of mm sources (serps16/serps18 and serps32/serps33) showed a more complex morphology. The residuals after fitting a single 2D Gaussian were significant in these cases, so we subsequently carried out fits using two 2D Gaussians. Even though there is an improvement over the single Gaussian fit, a more individual analysis should be done since this model does not suit the emission properly. In these cases, serps16 and serps33 show more compact dust emission compared with the extended emission of their neighbors serps18 and serps32, respectively.  Extended low-level emission, the presence of multiple components, and/or non-symmetric morphologies are probably affecting the Gaussian fitting.  Higher-angular resolution observations and sophisticated modeling (beyond the scope of this paper) would help to make better fits.

It is possible that some continuum detections actually pertain to ``background contaminants'', such as extragalactic sources.  Extrapolating from the calculation in \citet{Mac15}, based on works by \citet{Hod13, Kar13, Ono14} that include deep ALMA surveys, we estimate that within our mapped region (where power is greater than 30\%) fewer than four extragalactic sources should appear as contaminants among the mm sources we detect here and study as candidate protostars, at about a $3\sigma$ detection level (1 mJy).   This is a small fraction of our sources, and should not significantly affect trends we report.

\subsection{Source classifications and morphologies} \label{sec:blobs}

Note that in Table \ref{tab:sources} we list mm sources with and without IR counterparts, as well as IR sources with no mm emission counterpart.  IR sources were identified and classified using the catalog by \citetalias{Dun15}; a full presentation of the IR sources and their comparison with mm source counterparts is presented in Appendix \ref{app:IR}, and here we summarize briefly.   Fifteen mm sources coincide with IR sources classified by \citetalias{Dun15} as Class 0/I, and three mm sources coincide with ``flat-spectrum'' (``F'') sources (see also column 4 in Table \ref{tab:sources}).  Thirty-four mm sources do not coincide with any source in the \citetalias{Dun15} catalog.  We do not detect mm-wavelength continuum emission coincident with any of the three Class II sources in the same mapped region (see Figure \ref{fig:map}b); this can be explained because in the later YSO phase the dusty envelope has dispersed and the SED peak shifts to shorter wavelength. 

In the following, we discuss mm source morphologies and their coincidence with previous mm continuum studies.  Specifically, \citet{Plu15a} observed  2.7 mm continuum with CARMA and a beamsize of 5\as, and \citet{Mau11} observed 1.2 mm continuum with IRAM 30m and a beamsize of 11\as.  

First, in Figure \ref{fig:map_zoom} (c) we show what we call the central ``ridge'' of sources.  The continuum emission reveals six compact mm sources (serps40, serps41, serps45, serps46, serps47, serps48) aligned in the northwest-southeast direction within more diffuse emission (and the two sources serps36, serps44 in close proximity).  Within the more diffuse emission reside a total of fourteen mm sources (including also seprs32, serps33, serps35, serps37, serps38, serps39), spanning a region of $10-15\as$\ (4000-6000 au).  These sources are at the location of SerpS-MM18 detected by \citet{Mau11}.  In previous CARMA observations, two peaks of continuum emission (pertaining to CARMA-6 and CARMA-7) were identified in this same region, most closely corresponding to serps33 and serps45 (the brightest mm source) of the present study.  All 14 of the sources we detect with ALMA reside within the $5\sigma$\ level emission detected with CARMA around CARMA-6/7. A detailed dynamical study of molecular gas tracers should better characterize the sources near the ridge region; for example, we can study envelope kinematics and outflows being driving by the different sources.  The first case study was that of the episodic outflow from CARMA-7 (serps45) by \citet{Plu15b}.

To the north (Figure \ref{fig:map_zoom} a) of this ridge we also detect several sources that were previously undistinguishable with CARMA.  Coincident with the source we previously identified as CARMA-3 \citep[SerpS-MM16, by][]{Mau11}, we now identify three sources (serps16, serps18, serps19).  The two peaks of mm emission in closest proximity (serps16, serps18) were fit with simultaneous 2D Gaussians, and the distance between the peaks is $\lesssim2\as$ ($\lesssim800$\ au), motivating a future study of binarity for these sources.  To the south of serps16/18, we now detect continuum emission (serps17) corresponding to an IR class 0/I source that was previously undetected with mm-wavelengths. The source serps29 to the east (of serps16/18) is a class 0/I source previously identified as CARMA-5, and it also appears to have some elongation that was not seen in CARMA observations.  Additionally, we find several previously unidentified sources (serps25, serps28, serps30, serps34) nearby ($\lesssim10\as$, or 4000 au), including continuum emission (serps26) coincident with a class 0/I IR source previously undetected with mm-wavelengths.

In the southwest of our map we detect serps12 and serps14 (Figure \ref{fig:map_zoom} d) coincident with CARMA-1. These are two class 0/I sources that may also be binary protostars due to their proximity (on the plane of the sky).  In the northwest of our map (Figure \ref{fig:map_zoom} b) we detect a weak source (serps7) within $\sim4\as$\ of SerpS-MM10, which \citet{Mau11} claimed is a multiple source with several condensations within their $11\as$\ beam.  Since this source lies near the edge of our map, we may have reduced sensitivity at that location and/or sources may lie just outside the map edge.  

Finally, we mention that the source CARMA-2 from \citet{Plu15a} is not detected with ALMA in 1.3 mm continuum, and since it was only marginally detected by CARMA then we suggest that it is not in fact a likely protostar.  Similarly we do not detect emission with ALMA near the source SerpS-MM15 from \citet{Mau11}.  That source is classified by \citet{Mau11} as Class 0, with a relatively (compared with other sources in that work) large ratio of $L_{\lambda>350\mu m}/L_{bol}$ and a note that it is ``emerging at Herschel wavelengths''; nonetheless the physical significance of the \citet{Mau11} detection of this source and the non-detection with ALMA is not understood.  We also note, for clarity, that CARMA-4 from \citet{Plu15a} is outside the field of view of our current ALMA map.   

\subsection{Source masses} \label{sec:masses}

Sub-millimeter continuum flux densities allow us to calculate the mass of the gas and dust of the disk and/or the envelope of a source.  We measured mass for each source according to the following equation:
\begin{equation}
M=\frac{D^2 S_\nu}{B_\nu(T_d) \kappa_\nu},
\end{equation}
where $D$ is distance to the source,  $S_\nu$ is the 1.3 mm continuum flux, and $B_\nu(T_d)$ is the Planck function given a dust temperature $T_d$.  This calculation assumes an optically thin model and a constant temperature for each source.  We used a dust opacity of $\kappa_\nu=0.1(\nu/1200\textrm{ GHz})^\beta$\ cm$^2$g$^{-1}$, with $\beta=1$, which corresponds to $\kappa_\nu=0.0188$\ cm$^2$g$^{-1}$ at $\lambda=1.3$\ mm \citep{Loo00}.  These values are consistent with our previous study of Serpens South \citep{Plu15a}; however we also note that \citet{Mau11}\ use the value $\kappa=0.01$\ cm$^2$g$^{-1}$\ for protostellar envelopes, following \citet{Oss94}, which would result in masses greater by about a factor of 2.  We assigned $T_d=35$\ K for sources with IR counterparts and $T_d=15$\ K for deeply embedded sources without IR counterparts \citep[following][]{Loo00}.  We note that using $T_d=15$\ K versus $T_d=35$\ K for a given source results in a factor of 3 greater mass in the case of the lower temperature; indeed \citet{Mau11} used an average value of  $T_d=16$\ K in their study .  Consequently, if we were to use a single temperature (e.g.~either $T_d=15$\ K or $T_d=35$\ K) for all sources, then the mass-ranked list of sources would be affected with respect to the list that we report here (we mention the mass-ranked list because it is important for the analysis of \S \ref{sec:mst}).  For reference, in our study, we report that the most massive source with an IR counterpart is the 20th most massive source in the cluster, but if we were to use a single temperature for all sources then the most massive source with an IR counterpart would be the 12th most massive source.  Given the uncertainties in these parameters, the mass calculation may have an error of a factor of a few, but moreover the uncertainties in $\kappa_\nu$\ and $T_d$\ may compensate for each other.

The masses we measured are in the range of 0.002 to 0.9 $\msun$\, and given in Table \ref{tab:sources}.  The lowest mass source that we detect without an IR counterpart has a mass of  0.005 $\msun$.  Hereafter we refer to these masses as ``mm mass'' or "source mass", interchangeably, to mean disk and/or envelope; these masses do not include the mass of the protostellar embryos (if any).  The equivalent mm mass that we can detect at a level of $5\sigma$ is $0.002\msun$ (2 M$_{jup}$) (which is also the mass of the lowest-mass source that we detect). We note again that the rms is about two times greater near the bright mm sources, so we may not be able to detect such low mass sources near the cluster center.

\subsection{Pre-stellar source stability}
Sources that have millimeter-continuum emission but no IR counterpart are candidates to be pre-stellar or deeply embedded protostellar cores \citep{1994MNRAS.268..276W}.  Alternatively, a source can be starless if it has no evidence for hosting a protostar \citep{2007prpl.conf...33W,2007prpl.conf...17D}.  To test whether a core is a collapsing pre-stellar core or a stable starless core, we compare the core with a Bonnor-Ebert isothermal sphere, inferring the critical density ($n_{crit}$) and maximum radius ($R_{max}$) for stability.  

For each source with no IR counterpart we calculated critical density as \citep[see also][]{Pad04,DeG16,Hue17}:

\begin{equation}
\left[\frac{n_{crit} }{cm^{-3} }\right]= 1.089\times10^4 \left[\frac{M_{BE} }{\msun}\right]^{-2} \left[\frac{T }{10K}\right]^{3} ,
\end{equation}

where we used T=15 K as the typical temperature, and we assumed the Bonnor Ebert mass $M_{BE}$ is the mm mass (Table \ref{tab:sources}).  Next, for a given $n_{crit}$ and mm mass, we measured the corresponding radius: 

\begin{equation}
R_{max} = \left(\frac{M_{BE}/\mu}{4/3\pi n_{crit}}\right)^{1/3},
\end{equation}
where $\mu=2.73 m_{H}$ . For each source we compared $R_{max}$\ with the observed radius $R_{obs}$.  We use $R_{obs} = (\theta_{maj}+\theta_{min})/4$, where $\theta_{maj}$\ and $\theta_{min}$\ are major and minor axes of the FWHM of the Gaussian fit (see Table \ref{tab:sources}), and in the case of unresolved cores we used the major axis of the beamsize as an upperlimit for $\theta_{maj}$ and $\theta_{min}$.  A core can be considered unstable, or collapsing, in the case that the ratio $n_{obs}/n_{crit}>1$ and $R_{obs}/R_{max}<1$ .  We present the results in Table \ref{tab:source_stability}.  We conclude that twenty of the mm sources with no IR counterpart are likely unstable or collapsing, having met the ``unstable'' criteria according to both density and radius (and excluding any unresolved sources that have inconclusive results, marked with ?? in the Table \ref{tab:source_stability}).  This corresponds to 70\% of the sample of mm sources with no IR counterpart (excluding those with inconclusive results).

Among these twenty sources, nine sources have ratios clearly indicating they are unstable cores or undergoing accretion. These sources are located toward the center of the cluster: at the central ridge (serps40, 41, 45, 46 and 47) or in the nearby vicinity (serps16, 18, 33 and 38). The other eleven sources meeting the unstable criteria have values closer to one. Given estimated uncertainty of $\sim30$\% for the ratios, the criteria for instability of these latter sources should be considered more tentative.

\section{Discussion} \label{sec:discussion}

\subsection{Source masses relative to cluster ``center''} \label{sec:center}

Qualitatively one can see that the higher mass, brighter continuum sources are forming in a concentrated region within the dense gas ridge of the larger cluster, shown in Figure \ref{fig:map} and in zoom in Figure \ref{fig:map_zoom} (a,c).  We reiterate that the dataset in hand is unique because we probe a range of masses over 2 orders of magnitude and down to low masses that have previously been undetected in the cluster setting.  We plot mm mass versus distance from cluster center for each protostar in Figure \ref{fig:massvdist} (a).  We use symbols in this plot consistent with those in Figures \ref{fig:map}-\ref{fig:map_zoom} in order to indicate evolutionary stage of each source in the case that there is an IR counterpart, or to indicate that the mm source has no IR counterpart.

The cluster center is defined as the intensity-weighted center of the \nhp\ emission, shown with a cross in Figure \ref{fig:map}.  A consequent nuance of the methodology presented in this sub-section is that defining a cluster center is not trivial when admitting that the cluster does not have a simple geometry (i.e.~ spherical).  Since our investigation seeks to understand the relation of source mass distribution and their nascent dense gas environment, then defining the cluster center based on \nhp\ (dense gas) emission seems most logical.  Another alternative would have been to assign the location of the highest-intensity continuum emission as the cluster center, which is at a distance of 4500 au from the adopted cluster center based on \nhp. Qualitatively the mass-distance trend holds in either case.  

The mm sources in Serpens South appear to decline in mass smoothly from the cluster center, with the most massive sources near the center.  A trend line is shown in Figure \ref{fig:massvdist} (a) to guide the eye.  Including all mm sources, we measure a median distance from the cluster center of 0.06 pc.  This is about one-third of the median offset of 0.17 pc that was measured by \citet{Cyg17} for the sources in the massive cluster G11.92-0.61.  We note that they measured distance relative to their most massive continuum source MM1, and if we measure distance relative to the most massive source in Serpens South, still the median distance is small. Probably the smaller median distance measured in Serpens South is due to the improved mass sensitivity and resolution in our observations, given the proximity of the region.  It is apparent (i.e.~Figure \ref{fig:map_zoom} c) that we identify a number of low-mass sources very close to the cluster center that may have been undetected if this cluster were at a farther distance.  There also may be a difference in the clustered formation scenario for the low-mass region Serpens South, with a smaller mass reservoir at the center compared with the high-mass case of  G11.92-0.61.

\subsection{Minimum Spanning Tree} \label{sec:mst}

A stellar cluster is considered mass segregated when the more massive stars are concentrated toward the ``center'' (or where the potential well is deeper). In a protostellar cluster, the corresponding segregation should be apparent among prestellar objects. Mass segregation is a natural outcome of the dynamical gravitational interactions between the cluster members if enough time is allowed, but it remains unknown if the cluster can be primordially mass segregated before dynamical interactions become important. In this subsection, we analyze mass segregation for the first time in a very young protostellar cluster, Serpens South.

The Minimum Spanning Tree (MST) method quantifies clustering, and we adopt the algorithm described in \citet{All09b} \citep[see also][]{2011MNRAS.416..541M}.  Most generally, quantifying the clustering will allow us and others in the future to compare with observations results for other comparable regions and more evolved or more massive clusters; simulations can also implement these parameters.  The MST method has the virtue of avoiding problems such as setting the cluster center\footnote{We note that an alternative method to MST is that of local stellar surface number density as a function of stellar mass, as described by \citet{2011MNRAS.416..541M,2012MNRAS.427..637P,2014MNRAS.438..620P}.  That method also does not depend on determining the cluster center. Further it can avoid the possible effect of outliers to skew the MST method.  Since no massive outliers are apparent in our dataset, we consider that the MST method is sufficient for our study.}  (as we commented previously, this can be challenging).

The MST measures the shortest path length connecting all points in a given sample, without including closed loops \citep{Kru56}. We compute it by means of the csgraph routine in the Python scipy toolkit \citep{Jon01}, using a mass-ranked list of all mm sources. \citet{All09b} defined the Mass Segregation Ratio ($\Lambda_{MSR}$) as
\begin{equation}
\Lambda_{MSR}=\frac{L_{norm}}{L_{massive}}\pm \frac{\sigma_{norm}}{L_{massive}},
\end{equation}
where L$_{massive}$ is the MST of the $N$\ most massive sources and L$_{norm}$ is the average pathlength of the MST of $N$\ random sources in the cluster. We take the average over 500 random sets of $N$ sources in the cluster to estimate L$_{norm}$ and its statistical deviation \citep{Gav17}. If $\Lambda_{MSR}>1$, the pathlength of the $N$ most massive sources is shorter than the average pathlength of any $N$ sources in the cluster, therefore the massive sources are concentrated relative to a random sample, and hence the cluster is mass segregated. In addition, the larger the $\Lambda_{MSR}$, the more concentrated the massive sources.

Figure \ref{fig:separations} (a) shows the result of the MST analysis of the mm sources detected in Serpens South.  In order to minimize the possibility for evolutionary states of the sources to affect the direct comparison of mm masses (see discussion in \S \ref{sec:ageseg}), we separate the results for the thirty-four mm sources with no IR counterpart -- which are likely the pre-stellar or most embedded (youngest) Class 0 sources -- and the 15 mm sources with IR counterparts that are classified as Class 0/I. In the case of the youngest sources (pre-stellar cores or early Class 0/I, those without an IR counterpart), the dust and gas masses should represent a substantial amount of the total mass of the system, while for the more evolved mm sources the estimated mm mass only accounts for the disk and remnants of the nascent dusty envelope.

The diagram displays the $\Lambda_{MSR}$ parameter obtained for source sets with an increasing number of mass-ranked members ($N_{MST}$). The most distinct features are: (i) a very large value of the mass segregation ratio for $N_{MST}=2$\ ($\Lambda_{MSR}\sim43$, outside the range of the plot), (ii) an approximately flat plateau with a mean value of $\langle\Lambda_{MSR}\rangle \sim3.7$\ for $3\leq N_{MST}\leq18$, (iii) a second shorter plateau between $19\leq N_{MST}\leq21$ at $\Lambda_{MSR}\sim1.5$, (iv) a decrease of $\Lambda_{MSR}$ to $\Lambda_{MSR}\sim1$ for $N_{MST}\gtrsim21$. These features indicate degrees of mass hierarchization. First, the two most massive mm sources with no IR counterpart are very close to each other compared with the typical separation of the cluster members. In second place, the 18 most massive pre-stellar candidate sources (mass $>0.04$\msun) are mass segregated, although not hierarchically (i.e., these 18 are distributed approximately homogeneously). A third stage in the mass segregation of the cluster involves three sources with masses $\sim0.03$\msun. These are not as spatially concentrated as the 18 most massive, and this is reflected in the diagram. Beyond this mass, the rest of the sources show a steady decrease in the degree of mass segregation, approaching $\Lambda_{MSR}\sim1$.  In the MST analysis of mm sources with IR counterparts that are classified as Class 0/I, we see no evidence for segregation based on envelope/disk (mm wavelength) mass, but a study invoking modeling to determine protostellar masses could show if the segregation holds at this stage.

\subsection{Segregation according to mass or age effects} \label{sec:ageseg}

The results of the MST analysis suggest that the youngest millimeter sources (pre-stellar candidates) in the Serpens South central protocluster are hierarchically mass segregated. However, as described in \S \ref{sec:masses}, the source masses we calculate (and subsequently use for a mass-ranked list to make the MST calculations, as in \S \ref{sec:mst}) correspond to the mass of the gas and dust of the disk and/or the envelope, not the protostar mass itself. Given that the Serpens South sample of mm sources include a range of evolutionary classes (from pre-stellar to flat-SED sources), the observed mass segregation could be biased by age effects as the envelopes evolve. In order to minimize this effect, we limited the MST analysis to only the youngest mm sources (i.e., those without IR counterpart). Constraining the sample in this way, we deal with sources in the pre-stellar and earliest Class 0 stages, that we presume are roughly coeval. The mm emission of a pre-stellar source comes mainly from the dust of the envelope and reflects almost all the mass of the system \citep[protostar plus disk plus envelope,][]{And00}.  On the other hand, the mm emission of the Class 0/I sources comes from the envelope and the proto-disk, which together, should have a mass of at least half of the protostellar mass \citep[e.g.,][]{1996A&A...311..858B,And00}. 

To quantify how much mass has been accreted by one of our early Class 0 candidate sources without IR counterpart, we can take a look into the molecular outflow activity in the region \citep{Plu15a, Plu15b}, and make an estimate of the amount of mass accreted. Probably the most prominent outflow in the cluster stems from serps45 (aka CARMA 7), whose first ejection is less than 5000 years old (the time it would take for an ejection found 4 arcminutes from its driving source, traveling at 100 km/s at a distance of 436 pc; see also Plunkett et al. 2015b). Assuming that during the Class 0 stage the accretion rate is between $10^{-5}$\ to $10^{-6}$\ \msun\ yr$^{-1}$ \citep{1996A&A...311..858B,And00}, the total mass accreted by serps45 would be $\left(5\times 10^{-6}\ \msun \textrm{yr}^{-1}\right) \times \left(5 \times 10^{3}\textrm{ yr }\right)\sim 0.025\ \msun$. For the other sources, we consider shorter ejections (let us consider a 1\am\ outflow lobe, or approximately the size of the central cluster), and the accreted mass would be $\sim0.005$\ \msun; for even shorter ejections of 30\as (or shorter, perhaps even those young outflows that are undetected), the accreted mass is $\sim0.0025\ \msun$. The latter is about half of the minimum mm mass measured in our sample of mm sources with no IR counterpart. Thus, taking into account that the measured mm masses could be off by up to about 0.025 \msun, the conclusions of mass segregation according to MST analysis for the youngest sources holds; the additional mass uncertainty due to a reasonable estimate of accretion may only shift slightly the value  $N_{MST}$ at which a break in $\Lambda_{MSR}$ is seen.  Nonetheless, in this analysis the data cannot entirely rule out small contributions that might be the result of age segregation.

We also performed the MST analysis for the fifteen mm sources with IR detections and classification of Class 0/I, likely more evolved than the former case of sources without IR detections. In this case, the difference in envelope mass between the pre-stellar to Class I phases may be approximately an order of magnitude \citep{1996A&A...311..858B}.  The MST analysis shows no significant trend of mass segregation for these mm sources.  An additional difference in envelope mass by a factor of about 6 occurs between Class I to Class II \citep{1994ApJ...420..837A}, yet we find no mm detections corresponding to the known Class II sources in this region. The (three) flat spectrum mm sources are at the transition from Class I to Class II, so here we can consider them to be slightly more evolved than the Class I sources (D15 show that the timescales of Class 0/I and Flat-spectrum YSOs are 0.5 and 0.3 Myr, respectively). Their envelope mass should have decreased by a factor less than 60 (and likely closer to a factor of 10, as in Class I sources) since the pre-stellar stage. Hence, for these sources, the mass of the protostellar system may be very different from that of the mm gas and dust mass estimate and therefore we did not include them in further analysis.

\subsection{Timescale for mass segregation}

Assuming that this segregation is not strongly affected by age differences among the sources, then the following question arises: is this mass segregation a primordial feature? To answer this, it is useful to estimate the dynamical segregation time for a YSO (which we use here as a protostar plus disk plus envelope) of a given mass. We follow the \citet{1969ApJ...158L.139S} prescription for the time scale for mass segregation, $t_{segregation}$, of a YSO of mass $M_{YSO}$\ in a cluster of $N$\ YSOs with average stellar mass $<m_{cluster}>$:
$$t_{segregation}\approx\frac{<m_{cluster}>}{M_{YSO}}\cdot\frac{N}{8\ln{N}}\cdot t_{crossing},$$
where the crossing time of the cluster is:
$$t_{crossing}\approx 2R_{hm}/v_{disp}.$$
Here, $R_{hm}$\ is the radius of the cluster containing half its total mass, and $v_{disp}$ is the dispersion velocity. From the velocity dispersion map of the \nhp emission, we derive $v_{disp}=0.47$\ \kms\ (see also Figure \ref{fig:massvdist} c and \S \ref{sec:densegas}). We estimate $R_{hm}\approx65\arcsec$ corresponding to $R_{hm}\approx0.14$~pc at the assumed distance. The resulting crossing time is $t_{crossing}=0.6$\ Myr. We adopt a typical value for the average stellar mass assuming a typical IMF to be $<m_{cluster}>=0.4$\ \msun \citep{All10}, and we estimate a number of cluster members of $\sim100$ considering sources in the catalogs based on infrared to X-ray observations \citep[D15,][see also Appendix \ref{app:IR}, \ref{app:Xray}]{Get17}, and the millimeter ALMA detections that we present here. Hence we derive a mass segregation time of $t_{segregation}=0.7/M_{YSO}$~Myr (i.e., it will take $7\times10^5$~yr to segregate a 1\msun\ YSO or $1.4\times10^6$~yr for a 0.5\msun\ YSO). For a cluster with 150 members, $t_{segregation}=2/M_{YSO}$~Myr is even larger. 

These time estimates are larger than the estimated age of the cluster.  While possible that some of the (few) flat-spectrum YSOs have an age of up to $\sim1$\ Myr (D15), the vast majority of sources in the dense cluster center (i.e.~Figure \ref{fig:map_zoom} a,c) are younger (classified as Class 0/I, or very likely young protostars yet unclassified by IR diagnostics).  Near the central ridge where we find the most massive mm sources, the sources likely have ages of $\lesssim0.46-0.72$\ Myr, which is the duration of Class 0+I reported by D15.  In other words, the sources that most contribute to our conclusion that more massive sources are concentrated toward the cluster center could not have mass segregation due to a dynamical origin, and therefore the segregation is probably primordial in origin. 

\subsection{Projected separations of sources} \label{sec:separations}

In Figure \ref{fig:separations} (b) we show the distribution of projected separations for unique pairs of millimeter sources.  Projected separations range from 500-73,000 au (0.002-0.35 pc).   A peak is seen at $\sim11,000$\ au when considering all mm sources, and at $\lesssim12,000$\ au (0.06 pc) for mm sources with no IR counterpart.  Considering the latter case, a slight majority ($\sim60\%$) of the unique separations are less than this distance,  with a break and then a stable, low level of unique separations at greater distances.  When considering all mm sources, a steady decline is seen longward of the peak at $\sim11,000$\ au.  A contributing factor to a peak at this distance may be that the Jean's length (assuming $T_k=20$\ K and $n(H_2)=10^5$\ cm$^{-3}$) is 0.08 pc \citep[following][]{2015MNRAS.453.3785P,2012sse..book.....K}. A plot similar to our Figure \ref{fig:separations} (b) is shown by  \citet[][see their Figure 5]{Cyg17} for the massive cluster G11.92-0.61, based on ALMA observations of 1 mm continuum emission in that region (spanning a region of 0.65 pc, with linear resolution of $\sim1650\times1150$\ au).  Considering all mm sources, they found two peaks in projected separation at $\sim0.1$\ pc and $\sim0.3$\ pc, with the peak at $\sim0.1$\ pc being more prominent.  While our peak appears at a slightly smaller distance than their prominent peak, this is also likely due to the difference in spatial resolution of the observations.  We also see a tentative peak in the bin at 0.3 pc when considering all mm sources (consistent with their analysis), which may be a coincidence, but merits further investigation in more clusters.  The interpretation of the projected separation peaks by \citet{Cyg17} suggests competitive accretion based on the claim that higher-mass sources require a cluster environment (i.e.~ small projected separations) for their formation. Our results are also consistent with a similar cluster scenario, now for a lower-mass cluster.

\subsection{Millimeter sources and the dense gas environment} \label{sec:densegas}

\begin{figure*}[!ht]
\includegraphics[width=0.5\textwidth,angle=0]{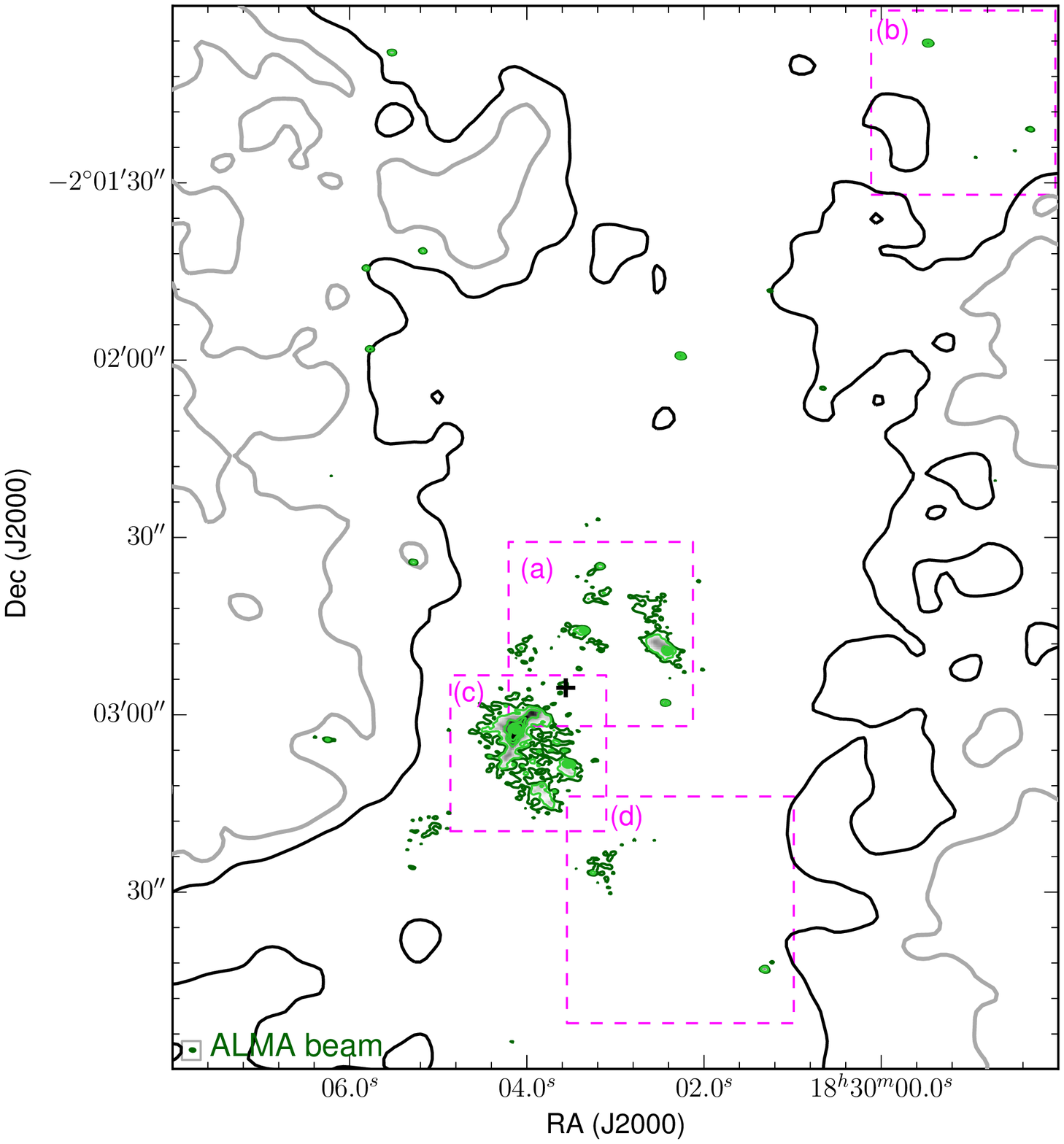}
\includegraphics[width=0.5\textwidth,angle=0]{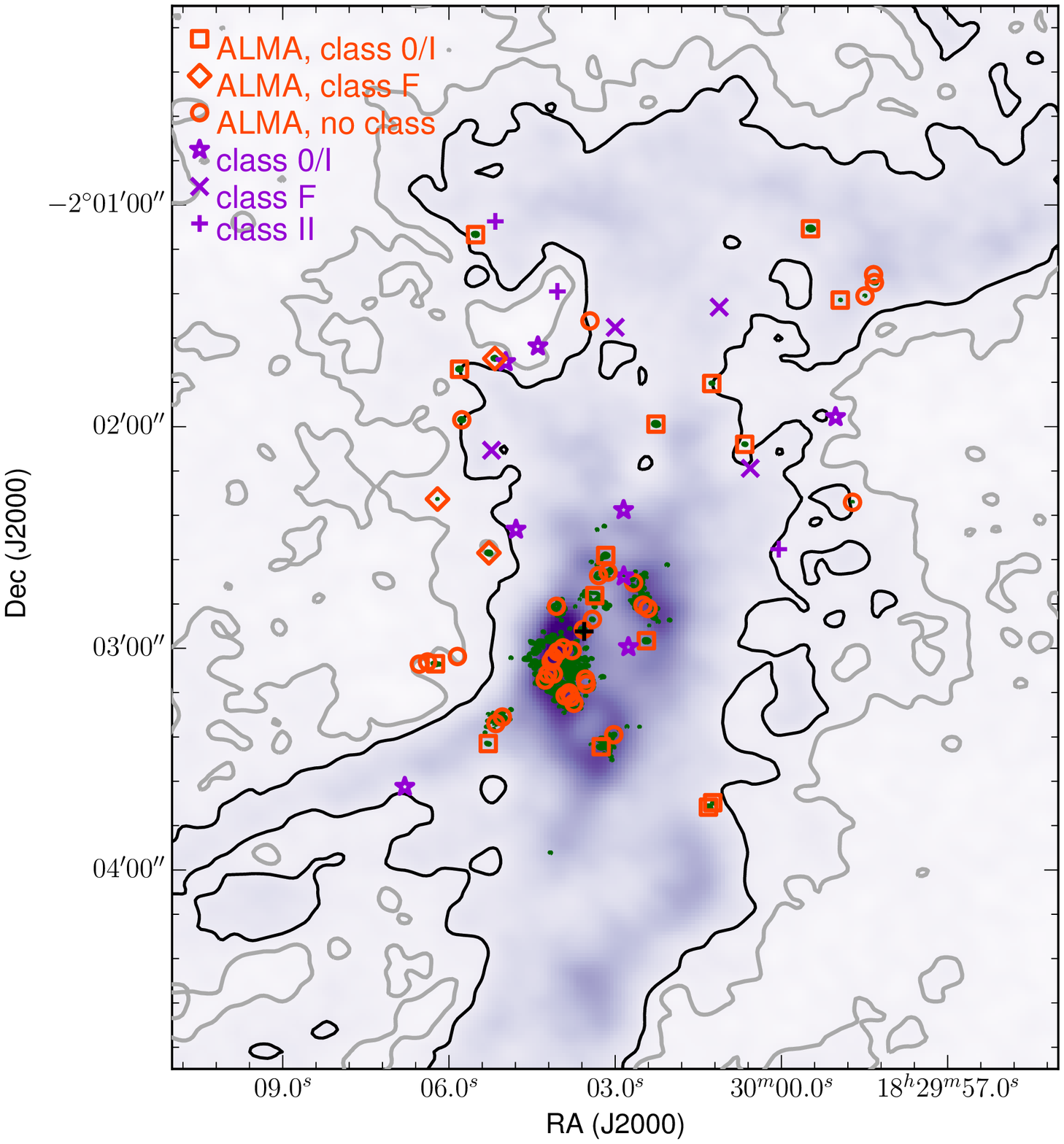}
\caption{(left) ALMA 1.3 mm continuum emission, shown in green contours with levels of 5$\sigma$ (dark green), 10$\sigma$,  30$\sigma$,  50$\sigma$,  70$\sigma$ (and thereafter increasing by increments of  50$\sigma$, where $\sigma=0.3$\ mJy bm$^{-1}$\ corresponds to rms noise in a signal-free region).  Gray and black contours show \nhp\ emission at levels of 5$\sigma$\ and 10$\sigma$, respectively.  Dashed magenta boxes outline the regions shown in ``zoom'' in Figure \ref{fig:map_zoom}. (right) \nhp\ emission is shown with grayscale (as well as contours as in left panel), zoomed out slightly from the region we mapped with ALMA in order to better display the dense gas filamentary structure of the region. Green contour shows lowest level (5$\sigma$) mm continuum emission as shown in left panel.  Symbols indicate all identified sources in the region.  Red (square, diamond, or circle) indicate a source detected with ALMA.  Purple (star, cross, or plus) indicate sources identified by \citetalias{Dun15}, but not detected with ALMA.  In this display it is evident that the YSOs are forming in the dense filamentary region, with no mm sources detected in a region where \nhp\ emission is lower than 5$\sigma$.   }
\label{fig:map}
\end{figure*}

Since the basic picture of star formation asserts that protostars form from the collapse of dense material \citep{Shu87}, then intuitively we should expect to observe protostars coincident with dense gas tracers.  In Figure \ref{fig:map} one can see that all sources reside within the $5\sigma$\ contour of the \nhp, and all but 4 sources reside within or touching the $10\sigma$\ contour of the \nhp.  The basic picture of star formation in dense gas holds; we suggest that the coincidence of mm sources and the dense gas is further evidence for a young age of the cluster, as protostellar sources have not had time to disperse from and are still controlled by their nascent dense gas \citep{Lad03,2010ARA&A..48..431P,Kru14,Stu16,Stu17}.  Is mass of the protostar also correlated with the dense gas?  Figure \ref{fig:massvdist} (b) shows mm mass as a function of \nhp\ integrated intensity at the same location (along the line of sight).  While we do not have the angular resolution in the \nhp\ data to actually resolve the same cores and measure their flux density, indeed higher mass sources are forming where the \nhp\ intensity is higher.  

We also find a trend between mm mass and \nhp\ velocity dispersion (as measured along the line of sight), shown in Figure \ref{fig:massvdist} (c).  A correlation of mass and velocity dispersion \citep[e.g., see][]{2014prpl.conf....3D} is consistent if gravitational infall and/or feedback via outflows dominates the dynamics in the cluster (on the scales observed here).  This is plausible given that the star formation in this cluster is recent and ongoing, and a high number density of young (Class 0/I) sources are clustered at the center where the protostar fraction reaches 91\% -- precisely the cases where infall and/or outflows are expected to be active.

\section{Summary} \label{sec:summary}

We presented an ALMA 1.3 mm continuum map of the protostellar cluster Serpens South.  According to previous IR observations, the cluster is reported to have a high protostar fraction, suggesting ongoing and active star formation. We identified 52 mm sources in our mapped region of $3\arcmin \times 2\arcmin$\ (equivalent to $0.38\times 0.25$\ pc).  These sources are protostellar candidates, given that the continuum emission corresponds to dust of the disk and/or the envelope.  In cases that the sources correspond with IR sources identified in the catalog by \citet{Dun15}, we indicate the evolutionary class of the source.  Eighteen mm sources coincide with IR sources classified by \citet{Dun15} as Class 0/I or ``flat-spectrum'' sources.  Thirty-four mm sources do not coincide with any source in the \citet{Dun15} catalog. Many of the latter case actually reside in the central ridge region where extinction negated the possibility for detection and classification at IR wavelengths, and therefore the ALMA data are powerful for identifying deeply embedded sources here.  

We measured mm source masses in the range of 0.002 to 0.9 $\msun$\ (see Table \ref{tab:sources}).  By analyzing the source masses and the spatial distribution within the cluster, we present evidence for mass segregation, specifically among the youngest mm sources (those with no IR counterpart).  Moreover, given indicators for the youth of the region, we suggest that the segregation is primordial.  Characteristic masses and distances apparent in the methods of Minimum Spanning Tree and projected separation (Figure \ref{fig:separations}) should provide parameters to compare with simulations and observations of other clusters. 

We studied in more detail the mm sources with no IR counterparts, considering these to be candidate pre-stellar or deeply embedded protostellar cores, or alternatively starless cores.  We analyzed the stability of the cores by comparing the critical density and the observed density.  About 60\% of the sample with no IR counterpart appear to be unstable or collapsing.

In addition to analysis of our ALMA observations, we utilize observational data from the literature for the same region.  We show that the mm source positions correspond with the dense gas traced by \nhp\, and mm mass correlates with \nhp\ intensity and velocity dispersion.  We compare (in the Appendix) the positions of mm sources in our study and corresponding detections using observational data at other wavelengths found in the literature.  These included (primarily) observations from \textit{Spitzer} and 2MASS \citep{Dun15}; Chandra \citep{Get17}; Herschel \citep{Kon15}; and VLA \citep{Ker16}.  We show that a multi-wavelength approach is important to clarify the evolutionary state of these sources.

\begin{acknowledgements}

We acknowledge the constructive suggestions by an anonymous referee that greatly enhanced the analysis presented in this paper. ALP thanks Amy Stutz, Claudia Cyganowski, and Stella Offner for insightful discussions of the interpretation of these data.  ALP obtained these data while hosted at Universidad de Chile with funding from Fulbright Chile.  HGA gratefully acknowledges support from the NSF under grant AST-1311825. GB is supported by the Spanish MINECO grant AYA2014-57369-C3.

This paper makes use of the following ALMA data: ADS/JAO.ALMA\#2012.1.00769.S. ALMA is a partnership of ESO (representing its member states), NSF (USA) and NINS (Japan), together with NRC (Canada), MOST and ASIAA (Taiwan), and KASI (Republic of Korea), in cooperation with the Republic of Chile. The Joint ALMA Observatory is operated by ESO, AUI/NRAO and NAOJ.  Data reduction was begun during a visit at the NRAO North American ALMA Science Center (NAASC), with generous assistance from Jennifer Donovan Meyer. The National Radio Astronomy Observatory is a facility of the National Science Foundation operated under cooperative agreement by Associated Universities, Inc.  

This research made use of Astropy, a community-developed core Python package for Astronomy \citep{2013A&A...558A..33A}, and the following SciPy \citep{Jon01} core packages: Pandas, Matplotlib, IPython, Numpy.%https://www.scipy.org/citing.html?]]

\end{acknowledgements}

\bibliographystyle{aa}

\begin{thebibliography}{80}
\expandafter\ifx\csname natexlab\endcsname\relax\def\natexlab#1{#1}\fi

\bibitem[{{Allison} {et~al.}(2009{\natexlab{a}}){Allison}, {Goodwin}, {Parker},
  {de Grijs}, {Portegies Zwart}, \& {Kouwenhoven}}]{All09a}
{Allison}, R.~J., {Goodwin}, S.~P., {Parker}, R.~J., {et~al.}
  2009{\natexlab{a}}, \apjl, 700, L99

\bibitem[{{Allison} {et~al.}(2010){Allison}, {Goodwin}, {Parker}, {Portegies
  Zwart}, \& {de Grijs}}]{All10}
{Allison}, R.~J., {Goodwin}, S.~P., {Parker}, R.~J., {Portegies Zwart}, S.~F.,
  \& {de Grijs}, R. 2010, \mnras, 407, 1098

\bibitem[{{Allison} {et~al.}(2009{\natexlab{b}}){Allison}, {Goodwin}, {Parker},
  {Portegies Zwart}, {de Grijs}, \& {Kouwenhoven}}]{All09b}
{Allison}, R.~J., {Goodwin}, S.~P., {Parker}, R.~J., {et~al.}
  2009{\natexlab{b}}, \mnras, 395, 1449

\bibitem[{{Andre} \& {Montmerle}(1994)}]{1994ApJ...420..837A}
{Andre}, P. \& {Montmerle}, T. 1994, \apj, 420, 837

\bibitem[{{Andr{\'e}} {et~al.}(2000){Andr{\'e}}, {Ward-Thompson}, \&
  {Barsony}}]{And00}
{Andr{\'e}}, P., {Ward-Thompson}, D., \& {Barsony}, M. 2000, Protostars and
  Planets IV, 59

\bibitem[{{Astropy Collaboration} {et~al.}(2013){Astropy Collaboration},
  {Robitaille}, {Tollerud}, {Greenfield}, {Droettboom}, {Bray}, {Aldcroft},
  {Davis}, {Ginsburg}, {Price-Whelan}, {Kerzendorf}, {Conley}, {Crighton},
  {Barbary}, {Muna}, {Ferguson}, {Grollier}, {Parikh}, {Nair}, {Unther},
  {Deil}, {Woillez}, {Conseil}, {Kramer}, {Turner}, {Singer}, {Fox}, {Weaver},
  {Zabalza}, {Edwards}, {Azalee Bostroem}, {Burke}, {Casey}, {Crawford},
  {Dencheva}, {Ely}, {Jenness}, {Labrie}, {Lim}, {Pierfederici}, {Pontzen},
  {Ptak}, {Refsdal}, {Servillat}, \& {Streicher}}]{2013A&A...558A..33A}
{Astropy Collaboration}, {Robitaille}, T.~P., {Tollerud}, E.~J., {et~al.} 2013,
  \aap, 558, A33

\bibitem[{{Bonnell} \& {Bate}(2006)}]{Bon06}
{Bonnell}, I.~A. \& {Bate}, M.~R. 2006, \mnras, 370, 488

\bibitem[{{Bonnell} \& {Davies}(1998)}]{Bon98}
{Bonnell}, I.~A. \& {Davies}, M.~B. 1998, \mnras, 295, 691

\bibitem[{{Bontemps} {et~al.}(1996){Bontemps}, {Andre}, {Terebey}, \&
  {Cabrit}}]{1996A&A...311..858B}
{Bontemps}, S., {Andre}, P., {Terebey}, S., \& {Cabrit}, S. 1996, \aap, 311,
  858

\bibitem[{{Bressert} {et~al.}(2010){Bressert}, {Bastian}, {Gutermuth},
  {Megeath}, {Allen}, {Evans}, {Rebull}, {Hatchell}, {Johnstone}, {Bourke},
  {Cieza}, {Harvey}, {Merin}, {Ray}, \& {Tothill}}]{Bre10}
{Bressert}, E., {Bastian}, N., {Gutermuth}, R., {et~al.} 2010, \mnras, 409, L54

\bibitem[{{Cheng} {et~al.}(2018){Cheng}, {Tan}, {Liu}, {Kong}, {Lim},
  {Andersen}, \& {Da Rio}}]{Che17}
{Cheng}, Y., {Tan}, J.~C., {Liu}, M., {et~al.} 2018, \apj, 853, 160

\bibitem[{{Cyganowski} {et~al.}(2017){Cyganowski}, {Brogan}, {Hunter}, {Smith},
  {Kruijssen}, {Bonnell}, \& {Zhang}}]{Cyg17}
{Cyganowski}, C.~J., {Brogan}, C.~L., {Hunter}, T.~R., {et~al.} 2017, \mnras,
  468, 3694

\bibitem[{{de Gregorio-Monsalvo} {et~al.}(2016){de Gregorio-Monsalvo},
  {Barrado}, {Bouy}, {Bayo}, {Palau}, {Morales-Calder{\'o}n}, {Hu{\'e}lamo},
  {Morata}, {Mer{\'{\i}}n}, \& {Eiroa}}]{DeG16}
{de Gregorio-Monsalvo}, I., {Barrado}, D., {Bouy}, H., {et~al.} 2016, \aap,
  590, A79

\bibitem[{{Delgado} {et~al.}(2013){Delgado}, {Djupvik}, {Costado}, \&
  {Alfaro}}]{2013MNRAS.435..429D}
{Delgado}, A.~J., {Djupvik}, A.~A., {Costado}, M.~T., \& {Alfaro}, E.~J. 2013,
  \mnras, 435, 429

\bibitem[{Di~Francesco {et~al.}(2007)Di~Francesco, Evans, Caselli, Myers,
  Shirley, Aikawa, \& Tafalla}]{2007prpl.conf...17D}
Di~Francesco, J., Evans, N. J.~I., Caselli, P., {et~al.} 2007, Protostars and
  Planets V, 17

\bibitem[{{Dobbs} {et~al.}(2014){Dobbs}, {Krumholz}, {Ballesteros-Paredes},
  {Bolatto}, {Fukui}, {Heyer}, {Low}, {Ostriker}, \&
  {V{\'a}zquez-Semadeni}}]{2014prpl.conf....3D}
{Dobbs}, C.~L., {Krumholz}, M.~R., {Ballesteros-Paredes}, J., {et~al.} 2014,
  Protostars and Planets VI, 3

\bibitem[{{Dunham} {et~al.}(2015){Dunham}, {Allen}, {Evans},
  {Broekhoven-Fiene}, {Cieza}, {Di Francesco}, {Gutermuth}, {Harvey},
  {Hatchell}, {Heiderman}, {Huard}, {Johnstone}, {Kirk}, {Matthews}, {Miller},
  {Peterson}, \& {Young}}]{Dun15}
{Dunham}, M.~M., {Allen}, L.~E., {Evans}, II, N.~J., {et~al.} 2015, \apjs, 220,
  11

\bibitem[{{Dzib} {et~al.}(2010){Dzib}, {Loinard}, {Mioduszewski}, {Boden},
  {Rodr{\'{\i}}guez}, \& {Torres}}]{Dzi10}
{Dzib}, S., {Loinard}, L., {Mioduszewski}, A.~J., {et~al.} 2010, \apj, 718, 610

\bibitem[{{Dzib} {et~al.}(2011){Dzib}, {Loinard}, {Mioduszewski}, {Boden},
  {Rodr{\'{\i}}guez}, \& {Torres}}]{Dzi11}
{Dzib}, S., {Loinard}, L., {Mioduszewski}, A.~J., {et~al.} 2011, in Revista
  Mexicana de Astronomia y Astrofisica, vol. 27, Vol.~40, Revista Mexicana de
  Astronomia y Astrofisica Conference Series, 231--232

\bibitem[{{Feigelson} \& {Montmerle}(1999)}]{Fei99}
{Feigelson}, E.~D. \& {Montmerle}, T. 1999, \araa, 37, 363

\bibitem[{{Feigelson} {et~al.}(2013){Feigelson}, {Townsley}, {Broos}, {Busk},
  {Getman}, {King}, {Kuhn}, {Naylor}, {Povich}, {Baddeley}, {Bate},
  {Indebetouw}, {Luhman}, {McCaughrean}, {Pittard}, {Pudritz}, {Sills}, {Song},
  \& {Wadsley}}]{Fei13}
{Feigelson}, E.~D., {Townsley}, L.~K., {Broos}, P.~S., {et~al.} 2013, \apjs,
  209, 26

\bibitem[{{Fern{\'a}ndez-L{\'o}pez} {et~al.}(2014){Fern{\'a}ndez-L{\'o}pez},
  {Arce}, {Looney}, {Mundy}, {Storm}, {Teuben}, {Lee}, {Segura-Cox}, {Isella},
  {Tobin}, {Rosolowsky}, {Plunkett}, {Kwon}, {Kauffmann}, {Ostriker}, {Tassis},
  {Shirley}, \& {Pound}}]{Fer14}
{Fern{\'a}ndez-L{\'o}pez}, M., {Arce}, H.~G., {Looney}, L., {et~al.} 2014,
  \apjl, 790, L19

\bibitem[{{Friesen} {et~al.}(2013){Friesen}, {Medeiros}, {Schnee}, {Bourke},
  {di Francesco}, {Gutermuth}, \& {Myers}}]{Fri13}
{Friesen}, R.~K., {Medeiros}, L., {Schnee}, S., {et~al.} 2013, \mnras, 436,
  1513

\bibitem[{{Gavagnin} {et~al.}(2017){Gavagnin}, {Bleuler}, {Rosdahl}, \&
  {Teyssier}}]{Gav17}
{Gavagnin}, E., {Bleuler}, A., {Rosdahl}, J., \& {Teyssier}, R. 2017, ArXiv
  e-prints [\eprint[arXiv]{1701.07982}]

\bibitem[{{Getman} {et~al.}(2017){Getman}, {Broos}, {Kuhn}, {Feigelson},
  {Richert}, {Ota}, {Bate}, \& {Garmire}}]{Get17}
{Getman}, K.~V., {Broos}, P.~S., {Kuhn}, M.~A., {et~al.} 2017, \apjs, 229, 28

\bibitem[{{Gutermuth} {et~al.}(2008){Gutermuth}, {Bourke}, {Allen}, {Myers},
  {Megeath}, {Matthews}, {J{\o}rgensen}, {Di Francesco}, {Ward-Thompson},
  {Huard}, {Brooke}, {Dunham}, {Cieza}, {Harvey}, \& {Chapman}}]{Gut08}
{Gutermuth}, R.~A., {Bourke}, T.~L., {Allen}, L.~E., {et~al.} 2008, \apjl, 673,
  L151

\bibitem[{{Gutermuth} {et~al.}(2009){Gutermuth}, {Megeath}, {Myers}, {Allen},
  {Pipher}, \& {Fazio}}]{Gut09}
{Gutermuth}, R.~A., {Megeath}, S.~T., {Myers}, P.~C., {et~al.} 2009, \apjs,
  184, 18

\bibitem[{{Hillenbrand} \& {Hartmann}(1998)}]{Hil98}
{Hillenbrand}, L.~A. \& {Hartmann}, L.~W. 1998, \apj, 492, 540

\bibitem[{{Hodge} {et~al.}(2013){Hodge}, {Karim}, {Smail}, {Swinbank},
  {Walter}, {Biggs}, {Ivison}, {Weiss}, {Alexander}, {Bertoldi}, {Brandt},
  {Chapman}, {Coppin}, {Cox}, {Danielson}, {Dannerbauer}, {De Breuck},
  {Decarli}, {Edge}, {Greve}, {Knudsen}, {Menten}, {Rix}, {Schinnerer},
  {Simpson}, {Wardlow}, \& {van der Werf}}]{Hod13}
{Hodge}, J.~A., {Karim}, A., {Smail}, I., {et~al.} 2013, \apj, 768, 91

\bibitem[{{Hu{\'e}lamo} {et~al.}(2017){Hu{\'e}lamo}, {de Gregorio-Monsalvo},
  {Palau}, {Barrado}, {Bayo}, {Ruiz}, {Zapata}, {Bouy}, {Morata},
  {Morales-Calder{\'o}n}, {Eiroa}, \& {M{\'e}nard}}]{Hue17}
{Hu{\'e}lamo}, N., {de Gregorio-Monsalvo}, I., {Palau}, A., {et~al.} 2017,
  \aap, 597, A17

\bibitem[{Jones {et~al.}(2001--)Jones, Oliphant, Peterson, {et~al.}}]{Jon01}
Jones, E., Oliphant, T., Peterson, P., {et~al.} 2001--, {SciPy}: Open source
  scientific tools for {Python}, [Online; accessed <today>]

\bibitem[{{Karim} {et~al.}(2013){Karim}, {Swinbank}, {Hodge}, {Smail},
  {Walter}, {Biggs}, {Simpson}, {Danielson}, {Alexander}, {Bertoldi}, {de
  Breuck}, {Chapman}, {Coppin}, {Dannerbauer}, {Edge}, {Greve}, {Ivison},
  {Knudsen}, {Menten}, {Schinnerer}, {Wardlow}, {Wei{\ss}}, \& {van der
  Werf}}]{Kar13}
{Karim}, A., {Swinbank}, A.~M., {Hodge}, J.~A., {et~al.} 2013, \mnras, 432, 2

\bibitem[{{Kern} {et~al.}(2016){Kern}, {Keown}, {Tobin}, {Mead}, \&
  {Gutermuth}}]{Ker16}
{Kern}, N.~S., {Keown}, J.~A., {Tobin}, J.~J., {Mead}, A., \& {Gutermuth},
  R.~A. 2016, \aj, 151, 42

\bibitem[{{Kippenhahn} {et~al.}(2012){Kippenhahn}, {Weigert}, \&
  {Weiss}}]{2012sse..book.....K}
{Kippenhahn}, R., {Weigert}, A., \& {Weiss}, A. 2012, {Stellar Structure and
  Evolution}

\bibitem[{{Kirk} \& {Myers}(2011)}]{2011ApJ...727...64K}
{Kirk}, H. \& {Myers}, P.~C. 2011, \apj, 727, 64

\bibitem[{{Kirk} {et~al.}(2013){Kirk}, {Myers}, {Bourke}, {Gutermuth},
  {Hedden}, \& {Wilson}}]{Kir13}
{Kirk}, H., {Myers}, P.~C., {Bourke}, T.~L., {et~al.} 2013, \apj, 766, 115

\bibitem[{{K{\"o}nyves} {et~al.}(2015){K{\"o}nyves}, {Andr{\'e}},
  {Men'shchikov}, {Palmeirim}, {Arzoumanian}, {Schneider}, {Roy}, {Didelon},
  {Maury}, {Shimajiri}, {Di Francesco}, {Bontemps}, {Peretto}, {Benedettini},
  {Bernard}, {Elia}, {Griffin}, {Hill}, {Kirk}, {Ladjelate}, {Marsh}, {Martin},
  {Motte}, {Nguy{\^e}n Luong}, {Pezzuto}, {Roussel}, {Rygl}, {Sadavoy},
  {Schisano}, {Spinoglio}, {Ward-Thompson}, \& {White}}]{Kon15}
{K{\"o}nyves}, V., {Andr{\'e}}, P., {Men'shchikov}, A., {et~al.} 2015, \aap,
  584, A91

\bibitem[{{Krumholz}(2014)}]{Kru14}
{Krumholz}, M.~R. 2014, \physrep, 539, 49

\bibitem[{{Kruskal}(1956)}]{Kru56}
{Kruskal}, J.~B. 1956, Proc.~Amer.~Math.~Soc., 7, 48

\bibitem[{{Kuhn} {et~al.}(2013){Kuhn}, {Povich}, {Luhman}, {Getman}, {Busk}, \&
  {Feigelson}}]{Kuh13}
{Kuhn}, M.~A., {Povich}, M.~S., {Luhman}, K.~L., {et~al.} 2013, \apjs, 209, 29

\bibitem[{{Lada} \& {Lada}(2003)}]{Lad03}
{Lada}, C.~J. \& {Lada}, E.~A. 2003, \araa, 41, 57

\bibitem[{{Lim} {et~al.}(2013){Lim}, {Chun}, {Sung}, {Park}, {Lee}, {Sohn},
  {Hur}, \& {Bessell}}]{Lim13}
{Lim}, B., {Chun}, M.-Y., {Sung}, H., {et~al.} 2013, \aj, 145, 46

\bibitem[{{Looney} {et~al.}(2000){Looney}, {Mundy}, \& {Welch}}]{Loo00}
{Looney}, L.~W., {Mundy}, L.~G., \& {Welch}, W.~J. 2000, \apj, 529, 477

\bibitem[{{MacGregor} {et~al.}(2015){MacGregor}, {Wilner}, {Andrews},
  {Lestrade}, \& {Maddison}}]{Mac15}
{MacGregor}, M.~A., {Wilner}, D.~J., {Andrews}, S.~M., {Lestrade}, J.-F., \&
  {Maddison}, S. 2015, \apj, 809, 47

\bibitem[{{Maschberger} \& {Clarke}(2011)}]{2011MNRAS.416..541M}
{Maschberger}, T. \& {Clarke}, C.~J. 2011, \mnras, 416, 541

\bibitem[{{Massi} {et~al.}(2003){Massi}, {Lorenzetti}, \& {Giannini}}]{Mas03}
{Massi}, F., {Lorenzetti}, D., \& {Giannini}, T. 2003, \aap, 399, 147

\bibitem[{{Massi} {et~al.}(2000){Massi}, {Lorenzetti}, {Giannini}, \&
  {Vitali}}]{Mas00}
{Massi}, F., {Lorenzetti}, D., {Giannini}, T., \& {Vitali}, F. 2000, \aap, 353,
  598

\bibitem[{{Maury} {et~al.}(2011){Maury}, {Andr{\'e}}, {Men'shchikov},
  {K{\"o}nyves}, \& {Bontemps}}]{Mau11}
{Maury}, A.~J., {Andr{\'e}}, P., {Men'shchikov}, A., {K{\"o}nyves}, V., \&
  {Bontemps}, S. 2011, \aap, 535, A77

\bibitem[{{Moeckel} \& {Bonnell}(2009)}]{2009MNRAS.396.1864M}
{Moeckel}, N. \& {Bonnell}, I.~A. 2009, \mnras, 396, 1864

\bibitem[{{Ono} {et~al.}(2014){Ono}, {Ouchi}, {Kurono}, \& {Momose}}]{Ono14}
{Ono}, Y., {Ouchi}, M., {Kurono}, Y., \& {Momose}, R. 2014, \apj, 795, 5

\bibitem[{{Ortiz-Le{\'o}n} {et~al.}(2017){Ortiz-Le{\'o}n}, {Dzib}, {Kounkel},
  {Loinard}, {Mioduszewski}, {Rodr{\'{\i}}guez}, {Torres}, {Pech}, {Rivera},
  {Hartmann}, {Boden}, {Evans}, {Brice{\~n}o}, {Tobin}, \& {Galli}}]{Ort17}
{Ortiz-Le{\'o}n}, G.~N., {Dzib}, S.~A., {Kounkel}, M.~A., {et~al.} 2017, \apj,
  834, 143

\bibitem[{{Ossenkopf} \& {Henning}(1994)}]{Oss94}
{Ossenkopf}, V. \& {Henning}, T. 1994, \aap, 291, 943

\bibitem[{{Padoan} \& {Nordlund}(2004)}]{Pad04}
{Padoan}, P. \& {Nordlund}, {\AA}. 2004, \apj, 617, 559

\bibitem[{{Palau} {et~al.}(2015){Palau}, {Ballesteros-Paredes},
  {V{\'a}zquez-Semadeni}, {S{\'a}nchez-Monge}, {Estalella}, {Fall}, {Zapata},
  {Camacho}, {G{\'o}mez}, {Naranjo-Romero}, {Busquet}, \&
  {Fontani}}]{2015MNRAS.453.3785P}
{Palau}, A., {Ballesteros-Paredes}, J., {V{\'a}zquez-Semadeni}, E., {et~al.}
  2015, \mnras, 453, 3785

\bibitem[{{Pang} {et~al.}(2013){Pang}, {Grebel}, {Allison}, {Goodwin},
  {Altmann}, {Harbeck}, {Moffat}, \& {Drissen}}]{Pan13}
{Pang}, X., {Grebel}, E.~K., {Allison}, R.~J., {et~al.} 2013, \apj, 764, 73

\bibitem[{{Parker} {et~al.}(2011){Parker}, {Bouvier}, {Goodwin}, {Moraux},
  {Allison}, {Guieu}, \& {G{\"u}del}}]{2011MNRAS.412.2489P}
{Parker}, R.~J., {Bouvier}, J., {Goodwin}, S.~P., {et~al.} 2011, \mnras, 412,
  2489

\bibitem[{{Parker} \& {Dale}(2013)}]{2013MNRAS.432..986P}
{Parker}, R.~J. \& {Dale}, J.~E. 2013, \mnras, 432, 986

\bibitem[{{Parker} {et~al.}(2012){Parker}, {Maschberger}, \& {Alves de
  Oliveira}}]{2012MNRAS.426.3079P}
{Parker}, R.~J., {Maschberger}, T., \& {Alves de Oliveira}, C. 2012, \mnras,
  426, 3079

\bibitem[{{Parker} \& {Meyer}(2012)}]{2012MNRAS.427..637P}
{Parker}, R.~J. \& {Meyer}, M.~R. 2012, \mnras, 427, 637

\bibitem[{{Parker} {et~al.}(2014){Parker}, {Wright}, {Goodwin}, \&
  {Meyer}}]{2014MNRAS.438..620P}
{Parker}, R.~J., {Wright}, N.~J., {Goodwin}, S.~P., \& {Meyer}, M.~R. 2014,
  \mnras, 438, 620

\bibitem[{{Plunkett} {et~al.}(2015{\natexlab{a}}){Plunkett}, {Arce}, {Corder},
  {Dunham}, {Garay}, \& {Mardones}}]{Plu15a}
{Plunkett}, A.~L., {Arce}, H.~G., {Corder}, S.~A., {et~al.} 2015{\natexlab{a}},
  \apj, 803, 22

\bibitem[{{Plunkett} {et~al.}(2015{\natexlab{b}}){Plunkett}, {Arce},
  {Mardones}, {van Dokkum}, {Dunham}, {Fern{\'a}ndez-L{\'o}pez}, {Gallardo}, \&
  {Corder}}]{Plu15b}
{Plunkett}, A.~L., {Arce}, H.~G., {Mardones}, D., {et~al.} 2015{\natexlab{b}},
  \nat, 527, 70

\bibitem[{{Portegies Zwart} {et~al.}(2010){Portegies Zwart}, {McMillan}, \&
  {Gieles}}]{2010ARA&A..48..431P}
{Portegies Zwart}, S.~F., {McMillan}, S.~L.~W., \& {Gieles}, M. 2010, \araa,
  48, 431

\bibitem[{{Sana} {et~al.}(2010){Sana}, {Momany}, {Gieles}, {Carraro},
  {Beletsky}, {Ivanov}, {de Silva}, \& {James}}]{San10}
{Sana}, H., {Momany}, Y., {Gieles}, M., {et~al.} 2010, \aap, 515, A26

\bibitem[{{Sault} {et~al.}(1995){Sault}, {Teuben}, \& {Wright}}]{Sau95}
{Sault}, R.~J., {Teuben}, P.~J., \& {Wright}, M.~C.~H. 1995, in Astronomical
  Society of the Pacific Conference Series, Vol.~77, Astronomical Data Analysis
  Software and Systems IV, ed. {R.~A.~Shaw, H.~E.~Payne, \& J.~J.~E.~Hayes},
  433--+

\bibitem[{{Schmeja} {et~al.}(2008){Schmeja}, {Kumar}, \&
  {Ferreira}}]{2008MNRAS.389.1209S}
{Schmeja}, S., {Kumar}, M.~S.~N., \& {Ferreira}, B. 2008, \mnras, 389, 1209

\bibitem[{{Shu} {et~al.}(1987){Shu}, {Adams}, \& {Lizano}}]{Shu87}
{Shu}, F.~H., {Adams}, F.~C., \& {Lizano}, S. 1987, ARAA, 25, 23

\bibitem[{{Spitzer}(1969)}]{1969ApJ...158L.139S}
{Spitzer}, Jr., L. 1969, \apjl, 158, L139

\bibitem[{{Storm} {et~al.}(2014){Storm}, {Mundy}, {Fern{\'a}ndez-L{\'o}pez},
  {Lee}, {Looney}, {Teuben}, {Rosolowsky}, {Arce}, {Ostriker}, {Segura-Cox},
  {Pound}, {Salter}, {Volgenau}, {Shirley}, {Chen}, {Gong}, {Plunkett},
  {Tobin}, {Kwon}, {Isella}, {Kauffmann}, {Tassis}, {Crutcher}, {Gammie}, \&
  {Testi}}]{Sto14}
{Storm}, S., {Mundy}, L.~G., {Fern{\'a}ndez-L{\'o}pez}, M., {et~al.} 2014,
  \apj, 794, 165

\bibitem[{{Strai{\v z}ys} {et~al.}(1996){Strai{\v z}ys}, {{\v C}ernis}, \&
  {Barta{\v s}i{\= u}t{\.e}}}]{Str96}
{Strai{\v z}ys}, V., {{\v C}ernis}, K., \& {Barta{\v s}i{\= u}t{\.e}}, S. 1996,
  Baltic Astronomy, 5, 125

\bibitem[{{Strai{\v z}ys} {et~al.}(2003){Strai{\v z}ys}, {{\v C}ernis}, \&
  {Barta{\v s}i{\= u}t{\.e}}}]{Str03}
{Strai{\v z}ys}, V., {{\v C}ernis}, K., \& {Barta{\v s}i{\= u}t{\.e}}, S. 2003,
  \aap, 405, 585

\bibitem[{{Stutz}(2017)}]{Stu17}
{Stutz}, A.~M. 2017, ArXiv e-prints [\eprint[arXiv]{1705.05838}]

\bibitem[{{Stutz} \& {Gould}(2016)}]{Stu16}
{Stutz}, A.~M. \& {Gould}, A. 2016, \aap, 590, A2

\bibitem[{{Tanaka} {et~al.}(2013){Tanaka}, {Nakamura}, {Awazu}, {Shimajiri},
  {Sugitani}, {Onishi}, {Kawabe}, {Yoshida}, \& {Higuchi}}]{Tan13}
{Tanaka}, T., {Nakamura}, F., {Awazu}, Y., {et~al.} 2013, \apj, 778, 34

\bibitem[{{Testi} {et~al.}(1998){Testi}, {Palla}, \& {Natta}}]{Tes98b}
{Testi}, L., {Palla}, F., \& {Natta}, A. 1998, \aaps, 133, 81

\bibitem[{{Testi} {et~al.}(1997){Testi}, {Palla}, {Prusti}, {Natta}, \&
  {Maltagliati}}]{Tes97}
{Testi}, L., {Palla}, F., {Prusti}, T., {Natta}, A., \& {Maltagliati}, S. 1997,
  \aap, 320, 159

\bibitem[{Ward-Thompson {et~al.}(2007)Ward-Thompson, Andre, Crutcher,
  Johnstone, Onishi, \& Wilson}]{2007prpl.conf...33W}
Ward-Thompson, D., Andre, P., Crutcher, R., {et~al.} 2007, Protostars and
  Planets V, 33

\bibitem[{Ward-Thompson {et~al.}(1994)Ward-Thompson, Scott, Hills, \&
  Andre}]{1994MNRAS.268..276W}
Ward-Thompson, D., Scott, P.~F., Hills, R.~E., \& Andre, P. 1994, Monthly
  Notices of the Royal Astronomical Society, 268, 276

\bibitem[{{Wright} {et~al.}(2014){Wright}, {Parker}, {Goodwin}, \&
  {Drake}}]{2014MNRAS.438..639W}
{Wright}, N.~J., {Parker}, R.~J., {Goodwin}, S.~P., \& {Drake}, J.~J. 2014,
  \mnras, 438, 639

\bibitem[{{Zinnecker} {et~al.}(1993){Zinnecker}, {McCaughrean}, \&
  {Wilking}}]{Zin93}
{Zinnecker}, H., {McCaughrean}, M.~J., \& {Wilking}, B.~A. 1993, in Protostars
  and Planets III, ed. E.~H. {Levy} \& J.~I. {Lunine}, 429--495

\end{thebibliography}

\clearpage

\begin{figure*}[!ht]
\includegraphics[width=\textwidth,angle=0]{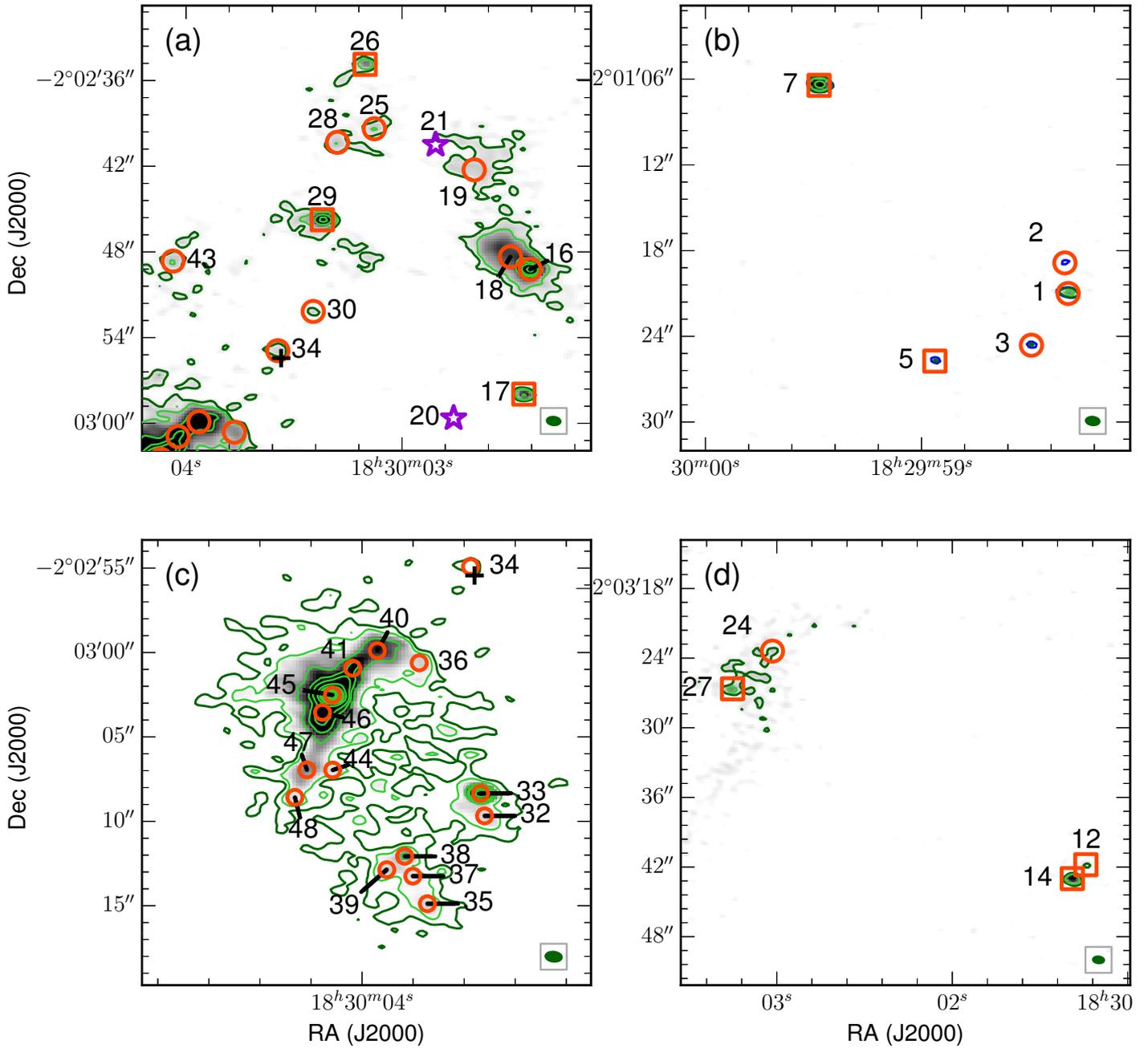}
\caption{Zoomed views into designated regions of map shown in Figure \ref{fig:map}, especially to show morphologies and source identifications in crowded and/or fragmented cases.  Contours are identical to Figure \ref{fig:map}, marking 5$\sigma$ (dark green), 10$\sigma$,  30$\sigma$,  50$\sigma$,  70$\sigma$ (and thereafter increasing by increments of  50$\sigma$).  Labels correspond to the source identification number in column 1 of Table \ref{tab:sources} (where sources are labeled as ``serps\#\#'').  }  
\label{fig:map_zoom}
\end{figure*}

\begin{figure*}[!ht]
\includegraphics[width=\textwidth]{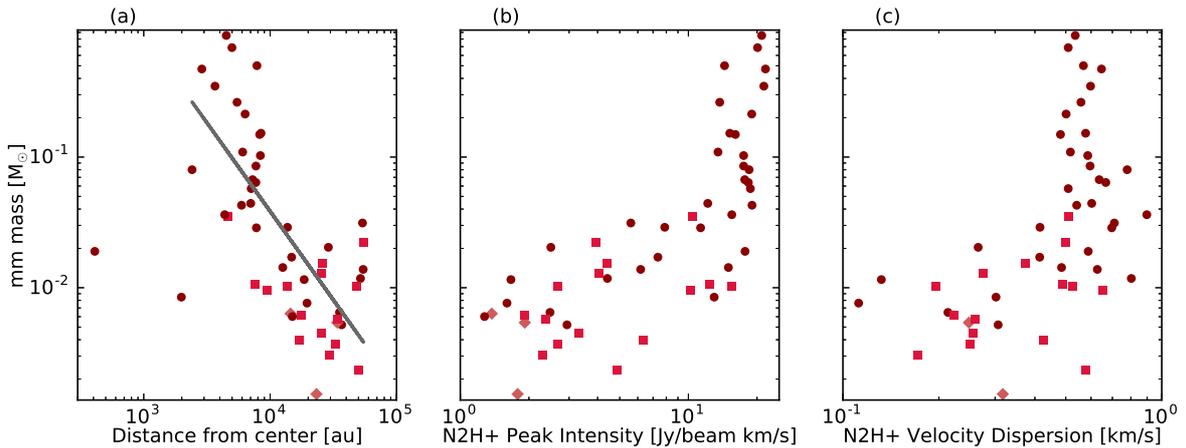}
\caption{ (a) Millimeter mass versus distance from ``center''.  We mark mm sources with no IR counterpart with dark red circles; Class 0/I protostar candidates with red squares; and flat-spectrum SED sources with light red diamonds.  The best fit power-law (gray line) has an index of $-1.4\pm0.2$.  We note that to perform this power-law fit, we removed the two sources with distances $d<2000$\ pc from the center (points in the lower left).  The ``center'' refers to the plus-sign in Figure \ref{fig:map}, and is discussed in \ref{sec:center}. (b) Mass versus \nhp\ integrated intensity. (c) Mass versus \nhp\ velocity dispersion.  Here ``mass'' refers to mass of dust and gas based on 1.3 mm continuum emission and the calculation described in \S \ref{sec:masses}.  \nhp\ integrated intensity and velocity dispersion are measured at each position given in Table \ref{tab:sources}.   }  
\label{fig:massvdist}
\end{figure*}

\begin{figure*}[!ht]
\includegraphics[width=\textwidth]{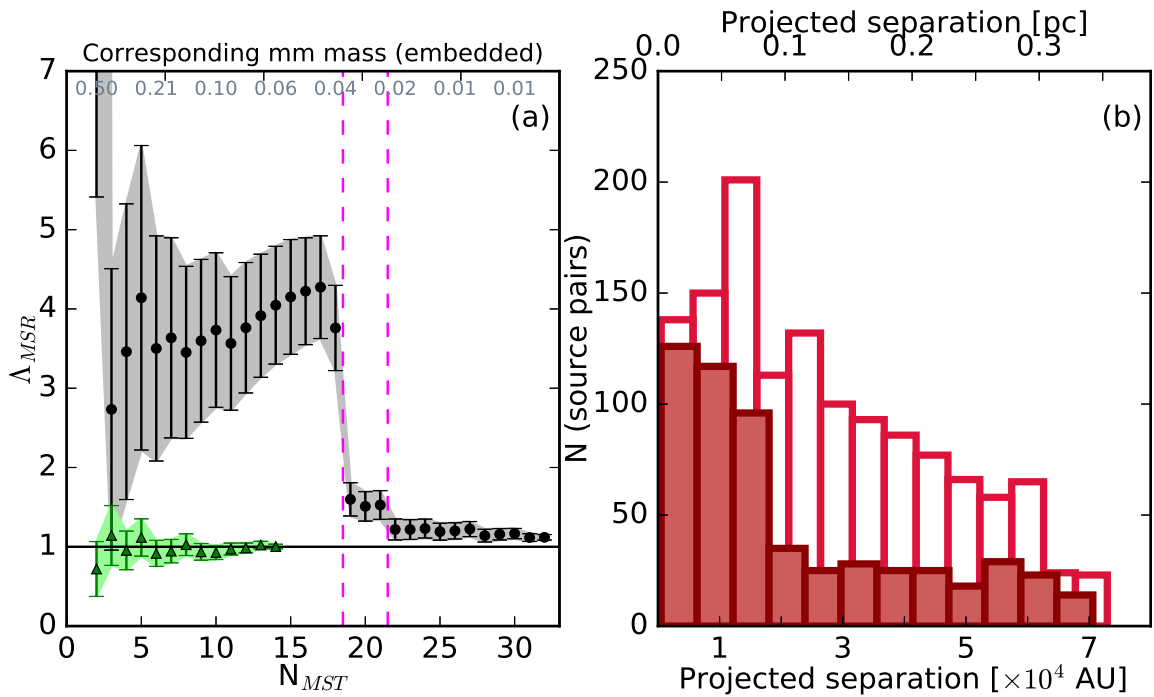}
\caption{ (a) Minimum spanning tree, as explained in \S \ref{sec:mst}, including with black, only mm sources with no IR counterpart (the most deeply embedded candidates); and with green, only mm sources with IR counterpart and classification of Class 0/I.  Vertical lines at $N=18$\ and $N=21$\ guide the eye to the principle and secondary breaks, suggesting that the 18 most massive sources are more clustered than random samples of 18 sources in the map, and there is a transition before evidence for clustering disappears for lower mass sources, $N\gtrsim21$.  Horizontal line at $\Lambda=1$\ marks the regime where sources show no evidence for mass segregation.  The upper x axis shows corresponding mm mass of the Nth source (considering only the embedded sources with no IR counterpart).  (b) Distribution of projected separations between unique pairs of millimeter sources (using Knuth binning scheme).  Dark red histogram (filled) shows only mm sources with no IR counterpart, while light red (unfilled) histogram shows all sources with mm detection in our map.} See \S \ref{sec:separations} for more discussion. 
\label{fig:separations}
\end{figure*}

\clearpage

\begin{table*}[hbt]
\caption[Sources]{Sources in mapped region} \label{tab:sources} 
\scriptsize
\setlength{\tabcolsep}{6pt}
\begin{center}
\begin{tabular}{lcccccccccccc}
\hline \hline
Label&
RA & Dec&
Class&
$\alpha$&
\multicolumn{2}{c}{FWHM \tablefootmark{a}}&
PA&
Peak&
Flux&
Mass&
D15 ID \tablefootmark{b}&
X | R \tablefootmark{c}\\
\cline{6-7}
&
hh:mm:ss & dd:mm:ss&
&
&
maj. (\as)&
min. (\as)&
(\dg)&
(mJy bm$^{-1}$)&
(mJy)&
(\msun) &
\\ 
\hline

serps1	&	 18:29:58.321  &-02:01:21.001  	&	  	&	  	&	 u 	&	 u 	&	 u 	&	10.23	&	10.34	&	0.031	&	  & X \\
serps2	&	 18:29:58.336  &-02:01:18.859  	&	  	&	  	&	 u 	&	 u 	&	 u 	&	3.62	&	4.56	&	0.014	&	  & - \\
serps3	&	 18:29:58.491  &-02:01:24.622  	&	  	&	  	&	0.51	&	0.25	&	81.0	&	3.16	&	3.89	&	0.012	&	  & X \\
serps4	&	 18:29:58.716  &-02:02:20.477  	&	  	&	  	&	 u 	&	 u 	&	 u 	&	2.52	&	2.14	&	0.006	&	  & -  \\
serps5	&	 18:29:58.938  &-02:01:25.735  	&	 0/I 	&	0.82	&	 u 	&	 u 	&	 u 	&	2.07	&	2.25	&	0.002	&	 J182958.9-020125 &X \\
serps6	&	 18:29:59.022  &-02:01:57.378  	&	 0/I 	&	0.35	&	  	&	  	&	  	&	  	&	  	&	  	&	 J182959.0-020157 & X\\
serps7	&	 18:29:59.474  &-02:01:06.43  	&	 0/I 	&	0.74	&	 u 	&	 u 	&	 u 	&	20.90	&	21.37	&	0.022	&	 J182959.4-020106 &X \\
serps8	&	 18:30:00.0466  &-02:02:33.1692  	&	 II 	&	-0.71	&	  	&	  	&	  	&	  	&	  	&	  	&	 J183000.0-020233 &X \\
serps9	&	 18:30:00.5609  &-02:02:11.274  	&	 F 	&	0.2	&	  	&	  	&	  	&	  	&	  	&	  	&	 J183000.5-020211 &X \\
serps10	&	 18:30:00.662  &-02:02:04.784  	&	 0/I 	&	0.45	&	 u 	&	 u 	&	 u 	&	2.73	&	2.93	&	0.003	&	 J183000.6-020204 &X \\
serps11	&	 18:30:01.122  &-02:01:27.6528  	&	 F 	&	-0.14	&	  	&	  	&	  	&	  	&	  	&	  	&	 J183001.1-020127 & X\\
serps12	&	 18:30:01.239  &-02:03:41.801  	&	 0/I 	&	1.12	&	0.90	&	0.64	&	-51.8	&	2.05	&	4.32	&	0.004	&	 J183001.3-020342 \tablefootmark{d}& X | 17\tablefootmark{e} \\
serps13	&	 18:30:01.257  &-02:01:48.32  	&	 0/I 	&	0.7	&	0.79	&	0.45	&	83.7	&	2.19	&	3.57	&	0.004	&	 J183001.2-020148 &X \\
serps14	&	 18:30:01.316  &-02:03:42.997  	&	 0/I 	&	1.12	&	0.33	&	0.26	&	11.0	&	10.38	&	12.33	&	0.013	&	 J183001.3-020342 \tablefootmark{d}& X | 17\tablefootmark{f} \\
serps15	&	 18:30:02.264  &-02:01:59.317  	&	 0/I 	&	0.55	&	 u 	&	 u 	&	 u 	&	14.26	&	14.74	&	0.015	&	 J183002.2-020159 & X \\
serps16	&	 18:30:02.407  &-02:02:49.219  	&	  	&	  	&	0.43\tablefootmark{i}	&	0.29	&	-20.6	&	38.22	&	49.18	&	0.149	&	  & -\\
serps17	&	 18:30:02.436  &-02:02:57.954  	&	 0/I 	&	0.48	&	 u 	&	 u 	&	 u 	&	10.68	&	10.22	&	0.011	&	 J183002.4-020257 & X\\
serps18	&	 18:30:02.498  &-02:02:48.337  	&	  	&	  	&	4.49\tablefootmark{i}	&	2.12	&	52.8	&	9.15	&	165.40	&	0.501	&	  & -\\
serps19	&	 18:30:02.666  &-02:02:42.254  	&	  	&	  	&	5.09	&	1.80	&	78.4	&	2.90	&	50.27	&	0.152	&	  &- \\
serps20	&	 18:30:02.7612  &-02:02:59.6364  	&	 0/I 	&	0.31	&	  	&	  	&	  	&	  	&	  	&	  	&	 J183002.7-020259 & X\\
serps21	&	 18:30:02.8447  &-02:02:40.4952  	&	 0/I 	&	0.97	&	  	&	  	&	  	&	  	&	  	&	  	&	 J183002.8-020240 &X \\
serps22	&	 18:30:02.8478  &-02:02:22.5456  	&	 0/I 	&	0.58	&	  	&	  	&	  	&	  	&	  	&	  	&	 J183002.8-020222 &X\tablefootmark{g} | 9\tablefootmark{h}  \\
serps23	&	 18:30:02.9986  &-02:01:33.1212  	&	 F 	&	0.05	&	  	&	  	&	  	&	  	&	  	&	  	&	 J183002.9-020133 & X\\
serps24	&	 18:30:03.025  &-02:03:23.399  	&	  	&	  	&	0.94	&	0.44	&	-52.2	&	2.47	&	4.72	&	0.014	&	  &- \\
serps25	&	 18:30:03.128  &-02:02:39.377  	&	  	&	  	&	1.44	&	0.74	&	-71.8	&	3.20	&	9.50	&	0.029	&	  &- \\
serps26	&	 18:30:03.171  &-02:02:34.87  	&	 0/I 	&	0.81	&	0.64	&	0.41	&	-67.0	&	6.13	&	9.22	&	0.010	&	 J183003.1-020234 & X | 10\tablefootmark{h}  \\
serps27	&	 18:30:03.252  &-02:03:26.647  	&	 0/I 	&	1.04	&	1.24	&	0.67	&	-83.4	&	3.97	&	9.88	&	0.010	&	 J183003.2-020326 & X | 16 \\
serps28	&	 18:30:03.299  &-02:02:40.327  	&	  	&	  	&	1.63	&	1.31	&	-51.4	&	3.00	&	14.60	&	0.044	&	  &- \\
serps29	&	 18:30:03.369  &-02:02:45.745  	&	 0/I 	&	0.84	&	0.99	&	0.54	&	-84.9	&	17.14	&	33.64	&	0.035	&	J183003.3-020245 &X|11 \\
serps30	&	 18:30:03.411  &-02:02:52.165  	&	  	&	  	&	 u 	&	 u 	&	 u 	&	2.69	&	2.80	&	0.008	&	  &- \\
serps31	&	 18:30:03.455  &-02:01:31.377  	&	  	&	  	&	 u 	&	 u 	&	 u 	&	1.64	&	1.72	&	0.005	&	  &- \\
serps32	&	 18:30:03.52  &-02:03:09.678  	&	  	&	  	&	2.26 \tablefootmark{i}	&	1.34	&	52.4	&	5.58	&	36.09	&	0.109	&	  & 13\tablefootmark{e}\\
serps33	&	 18:30:03.536  &-02:03:08.342  	&	  	&	  	&	0.47\tablefootmark{i}	&	0.32	&	74.8	&	68.55	&	86.78	&	0.263	&	  & 13\tablefootmark{f} \\
serps34	&	 18:30:03.575  &-02:02:54.913  	&	  	&	  	&	1.04	&	0.63	&	60.8	&	2.84	&	6.27	&	0.019	&	  & - \\
serps35	&	 18:30:03.744  &-02:03:14.879  	&	  	&	  	&	2.20	&	1.28	&	47.8	&	5.52	&	33.93	&	0.103	&	  & X | 14\tablefootmark{e}\\
serps36	&	 18:30:03.776  &-02:03:00.614  	&	  	&	  	&	1.73	&	1.14	&	44.1	&	5.73	&	26.44	&	0.080	&	  &- \\
serps37	&	 18:30:03.801  &-02:03:13.244  	&	  	&	  	&	2.17	&	1.24	&	71.8	&	4.88	&	28.22	&	0.085	&	  & 14\tablefootmark{e} \\
serps38	&	 18:30:03.833  &-02:03:12.071  	&	  	&	  	&	1.00	&	0.64	&	-67.2	&	10.10	&	22.19	&	0.067	&	  & 14\tablefootmark{f} \\
serps39	&	 18:30:03.903  &-02:03:12.861  	&	  	&	  	&	1.50	&	0.93	&	-62.8	&	5.91	&	21.11	&	0.064	&	  & 14\tablefootmark{e}\\
serps40	&	 18:30:03.94  &-02:02:59.886  	&	  	&	  	&	3.20	&	1.82	&	-84.4	&	13.76	&	156.00	&	0.472	&	  & - \\
serps41	&	 18:30:04.037  &-02:03:00.932  	&	  	&	  	&	2.60	&	1.08	&	-36.3	&	17.39	&	115.30	&	0.349	&	  & 12\tablefootmark{e} \\
serps42	&	 18:30:04.0433  &-02:01:23.4012  	&	 II 	&	-0.56	&	  	&	  	&	  	&	  	&	  	&	  	&	 J183004.0-020123 & X | 12\tablefootmark{e} \\
serps43	&	 18:30:04.06  &-02:02:48.639  	&	  	&	  	&	1.29	&	1.12	&	-42.0	&	3.28	&	11.97	&	0.036	&	  & - \\
serps44	&	 18:30:04.115  &-02:03:06.961  	&	  	&	  	&	1.55	&	0.54	&	-84.5	&	5.43	&	14.14	&	0.043	&	  & -  \\
serps45	&	 18:30:04.116  &-02:03:02.519  	&	  	&	  	&	0.77	&	0.68	&	-61.3	&	143.20	&	281.60	&	0.852	&	  & 12\tablefootmark{f} \\
serps46	&	 18:30:04.155  &-02:03:03.548  	&	  	&	  	&	2.60	&	1.55	&	-1.5	&	26.48	&	227.30	&	0.688	&	  & 12\tablefootmark{e} \\
serps47	&	 18:30:04.216  &-02:03:06.955  	&	  	&	  	&	2.99	&	1.19	&	-48.0	&	8.96	&	70.51	&	0.213	&	  & -  \\
serps48	&	 18:30:04.263  &-02:03:08.577  	&	  	&	  	&	1.46	&	1.05	&	-76.4	&	5.06	&	18.95	&	0.057	&	  & - \\
serps49	&	 18:30:04.3944  &-02:01:38.2332  	&	 0/I 	&	0.41	&	  	&	  	&	  	&	  	&	  	&	  	&	 J183004.3-020138 & X \\
serps50	&	 18:30:04.7894  &-02:02:27.8412  	&	 0/I 	&	0.65	&	  	&	  	&	  	&	  	&	  	&	  	&	 J183004.7-020227 &X \\
serps51	&	 18:30:04.9762  &-02:01:42.4848  	&	 0/I 	&	0.46	&	  	&	  	&	  	&	  	&	  	&	  	&	 J183004.9-020142 & - \\
serps52	&	 18:30:05.036  &-02:03:18.612  	&	  	&	  	&	2.59	&	0.66	&	81.3	&	2.21	&	9.59	&	0.029	&	  &-  \\
serps53	&	 18:30:05.153  &-02:03:20.367  	&	  	&	  	&	1.42	&	0.52	&	-56.5	&	2.15	&	5.66	&	0.017	&	  &-  \\
serps54	&	 18:30:05.1629  &-02:01:04.4364  	&	 II 	&	-0.32	&	  	&	  	&	  	&	  	&	  	&	  	&	 J183005.1-020104 &X \\
serps55	&	 18:30:05.171  &-02:01:41.561  	&	 F 	&	0.28	&	 u 	&	 u 	&	 u 	&	5.62	&	5.21	&	0.005	&	 J183005.1-020141 & X\\
serps56	&	 18:30:05.2308  &-02:02:06.3888  	&	 F 	&	0.17	&	  	&	  	&	  	&	  	&	  	&	  	&	 J183005.2-020206 & X\\
serps57	&	 18:30:05.278  &-02:02:34.176  	&	 F 	&	-0.3	&	0.56	&	0.21	&	78.0	&	4.97	&	6.11	&	0.006	&	 J183005.2-020234 & X\\
serps58	&	 18:30:05.291  &-02:03:25.795  	&	 0/I 	&	0.73	&	 u 	&	 u 	&	 u 	&	2.34	&	3.83	&	0.004	&	 J183005.2-020325 &X \\
serps59	&	 18:30:05.519  &-02:01:08.02  	&	 0/I 	&	0.39	&	 u 	&	 u 	&	 u 	&	10.21	&	9.96	&	0.010	&	 JJ183005.5-020107 &X \\
serps60	&	 18:30:05.768  &-02:01:58.145  	&	  	&	  	&	 u 	&	 u 	&	 u 	&	6.49	&	6.73	&	0.020	&	  & - \\
serps61	&	 18:30:05.809  &-02:01:44.455  	&	 0/I 	&	0.4	&	 u 	&	 u 	&	 u 	&	5.76	&	5.55	&	0.006	&	 JJ183005.8-020144 & X | 7 \\
serps62	&	 18:30:05.847  &-02:03:02.224  	&	  	&	  	&	0.47	&	0.27	&	63.6	&	1.61	&	1.99	&	0.006	&	  & -\\
serps63	&	 18:30:06.206  &-02:02:19.6  	&	 F 	&	-0.01	&	 u 	&	 u 	&	 u 	&	1.66	&	1.48	&	0.002	&	 JJ183006.2-020219 &X \\
serps64	&	 18:30:06.243  &-02:03:04.189  	&	 0/I 	&	0.79	&	1.03	&	0.17	&	81.9	&	3.92	&	5.96	&	0.006	&	 JJ183006.2-020304 &X \\
serps65	&	 18:30:06.386  &-02:03:03.713  	&	  	&	  	&	1.12	&	0.66	&	-74.3	&	1.62	&	3.81	&	0.012	&	  &- \\
serps66	&	 18:30:06.534  &-02:03:04.281  	&	  	&	  	&	 u 	&	 u 	&	 u 	&	1.56	&	2.52	&	0.008	&	  & -\\
serps67	&	 18:30:06.7968  &-02:03:37.4724  	&	 0/I 	&	0.32	&	  	&	  	&	  	&	  	&	  	&	  	&	 J183006.7-020337 & X\\
\hline
\hline 
\end{tabular}
\tablefoot{
\tablefoottext{a}{Where mm source was unresolved we mark with ``u''.  A blank indicates no corresponding mm source.}
\tablefoottext{b}{Source ID for matching source in \citetalias{Dun15} catalog.}
\tablefoottext{c}{``X'' indicates an X-ray source counterpart \citep{Get17}, while a number in the ``R'' column indicates the specific radio source counterpart labeled as VLA\#\# by \citet{Ker16}.}
\tablefoottext{d}{Two continuum sources are located within 2\as\ of the same IR source.}
\tablefoottext{e}{A radio source is located within 4\as\, but another mm source is closer to the location of the respective radio source.}
\tablefoottext{f}{This mm source is the best match to a radio source, but at least one other mm source is located within 4\as\ of the same radio source.}
\tablefoottext{g}{Tentative match between a mm source a X-ray source within 4\as. Note that the distance between this X-ray source and the radio source VLA9 is actually $\sim7\as$, with the mm source location in between. }
\tablefoottext{h}{Tentative match between a mm source and a radio source within 4\as.}
\tablefoottext{i}{Two 2D Gaussian fits were made for sources serps16/18 and serps32/33.  See \S \ref{sec:results} for discussion.}}
\end{center}
\end{table*}

\begin{table*}[hbt]
\caption[]{Source ``stability''} \label{tab:source_stability}
\scriptsize
\setlength{\tabcolsep}{6pt}
\begin{center}
\begin{tabular}{lccccrcl}
\hline \hline
&
\multicolumn{3}{c}{Density indicators }&
\multicolumn{3}{c}{Size indicators } \\
\cline{2-4}
\cline{5-7}
label&
$n_{obs}$&
$n_{obs}/n_{crit}$&
Unstable?&
$r_{obs}$\ \tablefootmark{a}&
$r_{obs}/r_{max}$&
Unstable?
\\
&
[cm$^{-3}$]&
[cm$^{-3}$]&
&
[au]&
[au]&
\\ \hline

serps1 & $>2.02e+08$ \tablefootmark{b} & 5.4 & True & $<209.7$ \tablefootmark{b} & 0.7 & True \\
serps2 & $>8.90e+07$ \tablefootmark{b} & 0.5 & ?? & $<209.7$ \tablefootmark{b} & 1.6 & False \\
serps3 & 2.63e+08 & 1.0 & False & 111.6 & 1.0 & False \\
serps4 & $>4.17e+07$ \tablefootmark{b} & 0.1 & ?? & $<209.7$ \tablefootmark{b} & 3.4 & ?? \\
serps16 & 5.77e+09 & 3480.2 & True & 92.9 & 0.1 & True \\
serps18 & 1.66e+07 & 113.1 & True & 978.8 & 0.2 & True \\
serps19 & 3.45e+06 & 2.2 & True & 1110.3 & 0.8 & True \\
serps24 & 5.15e+07 & 0.3 & False & 204.9 & 1.5 & False \\
serps25 & 2.87e+07 & 0.6 & False & 314.4 & 1.2 & False \\
serps28 & 3.08e+07 & 1.6 & True & 354.5 & 0.8 & True \\
serps30 & $>5.46e+07$ \tablefootmark{b} & 0.1 & ?? & $<209.7$ \tablefootmark{b} & 2.6 & ?? \\
serps31 & $>3.35e+07$ \tablefootmark{b} & 0.02 & ?? & $<209.7$ \tablefootmark{b} & 4.3 & ?? \\
serps32 & 2.83e+07 & 9.2 & True & 493.3 & 0.5 & True \\
serps33 & 7.49e+09 & 14056.5 & True & 102.9 & 0.04 & True \\
serps34 & 5.07e+07 & 0.5 & False & 226.5 & 1.3 & False \\
serps35 & 2.90e+07 & 8.3 & True & 478.9 & 0.5 & True \\
serps36 & 4.63e+07 & 8.1 & True & 377.1 & 0.5 & True \\
serps37 & 2.52e+07 & 5.0 & True & 472.4 & 0.6 & True \\
serps38 & 2.01e+08 & 24.6 & True & 218.2 & 0.3 & True \\
serps39 & 5.70e+07 & 6.3 & True & 326.6 & 0.5 & True \\
serps40 & 4.33e+07 & 262.8 & True & 696.9 & 0.2 & True \\
serps41 & 5.95e+07 & 197.2 & True & 566.8 & 0.2 & True \\
serps43 & 5.12e+07 & 1.8 & True & 280.1 & 0.8 & True \\
serps44 & 3.46e+07 & 1.7 & True & 337.5 & 0.8 & True \\
serps45 & 5.69e+09 & 112371.3 & True & 167.0 & 0.02 & True \\
serps46 & 1.18e+08 & 1516.4 & True & 566.1 & 0.1 & True \\
serps47 & 2.39e+07 & 29.6 & True & 652.5 & 0.3 & True \\
serps48 & 5.50e+07 & 4.9 & True & 318.7 & 0.6 & True \\
serps52 & 5.02e+06 & 0.1 & False & 564.2 & 2.1 & False \\
serps53 & 1.80e+07 & 0.1 & False & 309.1 & 1.9 & False \\
serps60 & $>1.31e+08$ \tablefootmark{b} & 1.5 & True & $<209.7$ \tablefootmark{b} & 1.1 & ?? \\
serps62 & 1.78e+08 & 0.2 & False & 101.6 & 1.8 & False \\
serps65 & 2.45e+07 & 0.1 & False & 244.6 & 2.2 & False \\
serps66 & $>4.91e+07$ \tablefootmark{b} & 0.1 & ?? & $<209.7$ \tablefootmark{b} & 2.9 & ?? \\
\hline
\hline 
\end{tabular}
\tablefoot{
\tablefoottext{a}{$r_{obs}$ corresponds to half of the mean of the major and minor axes of deconvolved fit (see Table \ref{tab:sources}), except in the cases where the source was unresolved.  For an unresolved source, $r_{obs}$\ corresponds to the major axis of the beamsize, and serves as an upper limit.}
\tablefoottext{b}{Where source size could not be deconvolved with beam, we use beamsize to determine lower limit on $n_{obs}$ and upper limit on $r_{obs}$.  In some cases this results in the unstable criteria to be inconclusive (marked with ``??'').} }
\end{center}
\end{table*}

\clearpage

\appendix
\section{Comparing with the literature}

\begin{figure}[!ht]
\includegraphics[width=\columnwidth,angle=0]{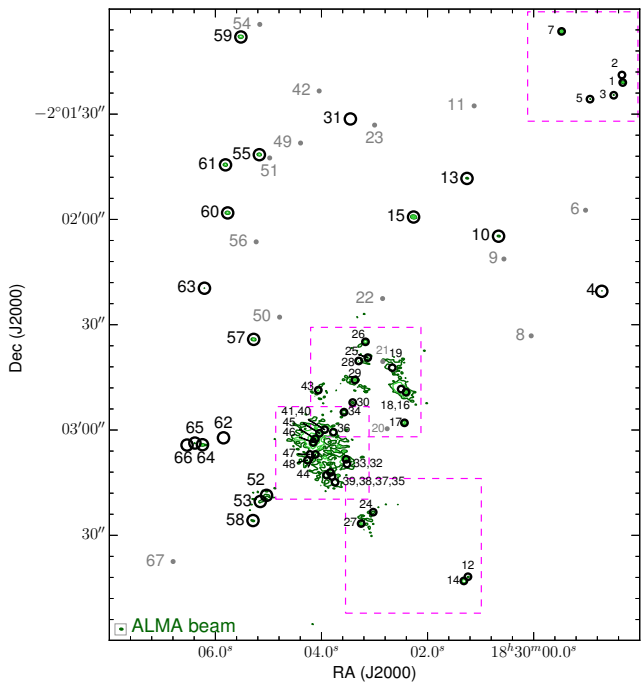}
\caption{Reference map with source labels, corresponding to the numbers in column 1 of Table \ref{tab:sources} (in the format ``serps\#\#'''). Contours and magenta boxes are the same as in Figure \ref{fig:map}, and boxes correspond to regions shown in Figure \ref{fig:map_zoom} in zoom.  We simplify the color and marker scheme from Figures \ref{fig:map}-\ref{fig:map_zoom}.  Here, black indicates a mm source, and gray indicates an IR source with no mm source counterpart.  Circle sizes of the markers are not significant, but are smaller in the crowded regions where a zoom version with labels is shown in Figure \ref{fig:map_zoom}.  }
\label{fig:map_with_labels}
\end{figure}

In this appendix we provide a qualitative review and comparison with sources identified in the literature corresponding to the same region that we mapped with ALMA.  The works referenced here utilize (primarily) observations from \textit{Spitzer} and 2MASS \citepalias{Dun15}; Chandra \citep{Get17}; Herschel \citep{Kon15}; and VLA \citep{Ker16}.  In Figure \ref{fig:map_with_labels} we provide a map (with observational data identical to that of Figure \ref{fig:map}) including source labels to guide the discussion. 

\subsection{Infrared (IR) catalog} \label{app:IR}

In Figures \ref{fig:map}-\ref{fig:map_zoom} and Table \ref{tab:sources} we showed the mm sources that are coincident (on plane of the sky) with IR sources from the catalog by \citetalias{Dun15}.  We also indicated the evolutionary class of the sources based on their spectral energy distribution (SED) using available data from \textit{Spitzer}\ and 2MASS between 2 $\mu$m and  24 $\mu$m \citepalias[see][for a full explanation of SED classifications]{Dun15}.  An object at the earliest stage of formation has an SED that peaks at longer wavelengths than an object in a later stage, and the corresponding slope of the IR SED also changes with time.

\begin{figure}[!ht]
\includegraphics[width=0.5\textwidth,angle=0]{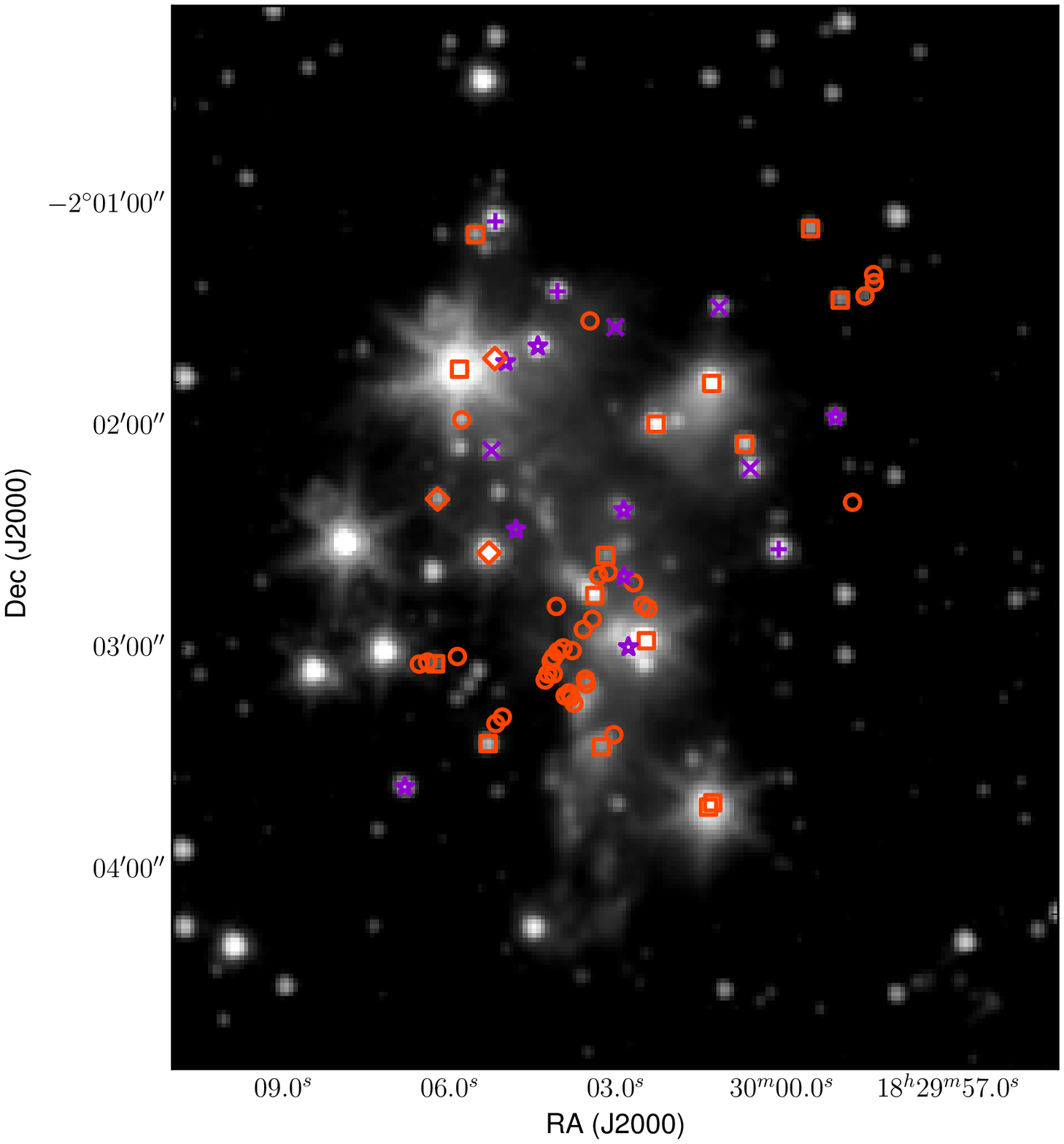}
\includegraphics[width=0.5\textwidth,angle=0]{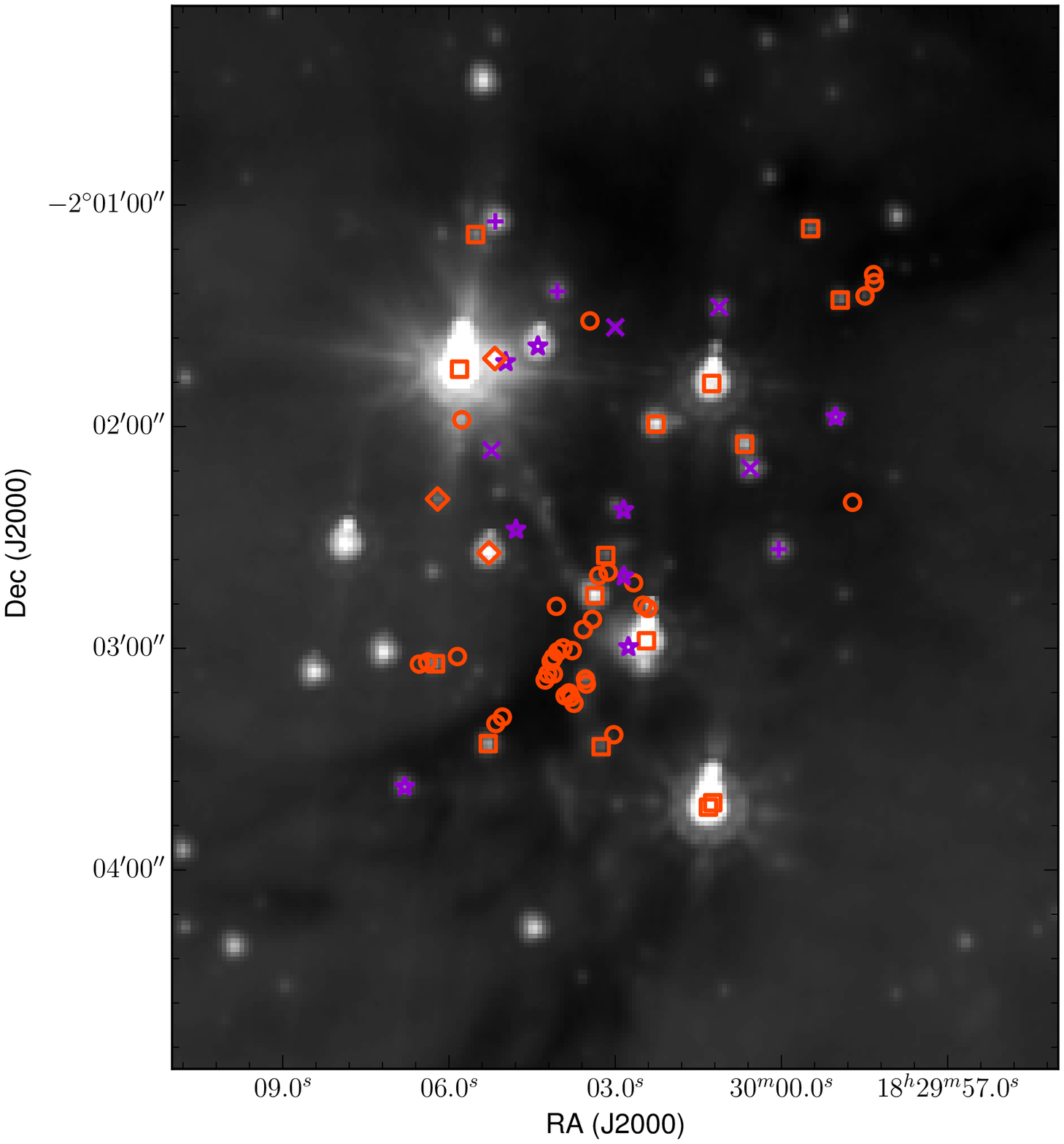}%updated 9/22/17

\caption{IRAC 2 (left) and IRAC 4 (right) maps with same source markers as Figure \ref{fig:map}. These maps can give a sense for sources where continuum emission correspond to a clear and isolated source as seen with \textit{Spitzer}, or where some confusion with nearby sources is possible.  Note that YSO identifications and classifications by \citetalias{Dun15} following the c2d method required detections in all four IRAC bands (3.6, 4.5, 5.8, 8.0 $\mu$m) and the MIPS 24 $\mu$m band.  }
\label{fig:IRAClabels}
\end{figure}

In Figure \ref{fig:IRAClabels} we show the IRAC 4.5 $\mu$m and 8.0 $\mu$m images (IRAC bands 2 and 4), and we mark the thirty-two YSOs from the \citetalias{Dun15} catalog that are located within the region we mapped with ALMA.  These 32 sources are classified as: 22 class 0/I; 7 flat-spectrum; 3 class II (see Table \ref{tab:sources} and Figure \ref{fig:map}).  One can clearly see the high protostar fraction (ratio of class 0/I protostars versus all YSOs) in the region that we mapped.  

Seventeen IR sources classified as YSOs are coincident within  2\arcsec\ of a mm source.  In the case of the IR source J183001.3-020342 (class 0/I), two mm sources (serps14 and serps12) appear coincident within 2\arcsec.  Considering the sources with both mm continuum detections and IR counterparts, 15 mm sources are classified as Class 0/I, and 3 as flat-spectrum, based on the spectral index from D15.  

There are 34 mm continuum sources with no IRAC counterpart within 2\arcsec. Expanding the match radius and inspecting closely by eye allows us to find some additional tentative matches.  Within 5\arcsec of J183006.2-020304 are located three more sources, in addition to serps64 that most closely matched.  Two sources are within 7\arcsec\ of J183005.2-020325, in addition to serps58. One source (serps24, see Figure \ref{fig:map_zoom}d) is located $\sim5$\arcsec\ from J183003.2-020326, and might be obscured by its bright IR emission.  Three mm sources in the northwest of our map are near ($\lesssim12$\arcsec) to serps5, and at least one of these appears to correlate with emission in IRAC2 and IRAC4 bands (Figure \ref{fig:IRAClabels}).  A possible IR source at the location of serps60 may be confused with the much brighter emission from the source J183005.8-020144 about 12\arcsec\ to the north, yet still appears to correspond to emission in IRAC2 and IRAC4 bands.  Hence ten mm sources may indeed have IR counterpart that could not be identified by \citetalias{Dun15} because they are clustered and/or faint.  In \citetalias{Dun15}, they also caution that their method is prone to missing these sources, because to be included in their catalog a source must have detections in all four IRAC bands (3.6-8.0 $\mu$m) as well as the first MIPS band (24 $\mu$m), with the MIPS observation having lower sensitivity and resolution \citep[see also][]{Gut09}.

serps31 has no apparent IR counterpart, but it appears situated between serps42 and serps23 in a region of diffuse IRAC2 emission.  If it is possibly a weak IR source, it might have been unidentified in the IR catalog.   Another mm source, serps4, does not correspond with IR emission and is relatively isolated.  Both of these sources are weak mm sources, and possibly they are not YSOs but instead background extragalactic sources.

Twenty-two mm sources without IR counterparts are located in the central cluster (see panels (a) and (c) of Figure \ref{fig:map_zoom}) where obscuration and blending probably negated the possibility for confident identification in IR images.  Still we consider these as deeply embedded protostellar candidates.

Fifteen sources classified as YSOs in the \citetalias{Dun15} catalog do not have a mm continuum counterpart (note that a few are near the edge of our mapped region, and if low mass may have been below our detection limit for those edge regions).  Among these are 8 class 0/I; 4 flat spectrum; and 3 class II.  \citetalias[][see their Figure 1]{Dun15} show that the contamination rate for sources with spectral index $\geq-0.7$ is close to zero.  All the sources coincident with our mapped region meet this criterion, hence we do not expect significant contamination according to that criterion.  Closer inspection of the corresponding SEDs is needed to better characterize classes and characteristics of these sources, but beyond the scope of the present work.

\subsection{X-ray and IR catalog} \label{app:Xray}

We compare our continuum sources with the SFiNCs (Star Formation in Nearby Clouds) catalog of \citet{Get17} that includes Chandra X-ray data for source selection and \textit{Spitzer} IR data for source classification.  X-ray surveys detect sources with magnetic reconnection flaring near the stellar or disk surface, and \citet{Get17} argue that these detections include disk-free stars that would be missed with IR emission, as well as Class I and II sources \citep[see also][]{Fei99,Fei13}.  

\begin{figure}[!ht]
\includegraphics[width=\columnwidth,angle=0]{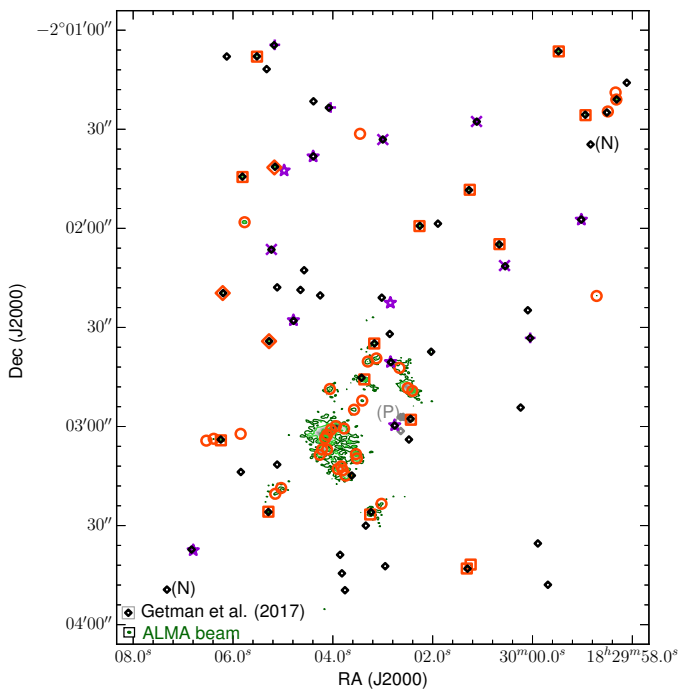}
\caption{ Locations of objects identified in the SFiNCS catalog by \citet{Get17} shown in diamonds. Black diamonds with no labels are classified as disk sources.  Two black diamonds with labels ``(N)'' are sources with no disk.  A small cluster of four gray diamonds labeled ``(P)''  are ``probable member'' sources.  We do not detect any mm sources coincident with ``no disk'' or ``probable member'' sources. Rather, all X-ray sources coincident with mm sources are classified as ``disk'' sources.  Continuum map and red/purple markers are the same as in Figure \ref{fig:map}. }
\label{fig:xray}
\end{figure}

Within the same region that we mapped with ALMA, \citet{Get17} identified 63 sources (Figure \ref{fig:xray}).  Twenty-one of the ALMA continuum detections coincide with the location of X-ray identified sources within the matching radius of 2\arcsec used by \citet{Get17,Kuh13}.  All but one of these sources (serps35 near the central cluster) match within $\lesssim1\arcsec$ (the ALMA beamsize).  We indicate these in Table \ref{tab:sources} (right column).

All of the ALMA mm sources coincident with SFiNCs sources have SFiNCs classification of ``DSK''.  Two mm sources in the northwest (serps1, serps3) and one mm source (serps35) near the central cluster that did not have counterparts identified in the IR catalog by D15 indeed have SFiNCs counterparts.  We show in Figure \ref{fig:xray} that in this region two X-ray identified sources are classified as ``NOD'' (``no disk''), and four sources are classified as ``PMB'' (``probable member''), yet do not have any mm continuum emission.

The thirty-one ALMA continuum detections that do not coincide with the location of an X-ray identified source fall into three categories: 21 are located in the central cluster (Figure \ref{fig:map_zoom}a, c) where obscuration and blending probably negated the possibility for an IR + X-ray catalog to confidently classify these sources; 8 are located within a few arcseconds of another continuum source that indeed coincides with an X-ray identified source; and 2 are isolated, weak continuum detections.  

The twenty-one sources in the central cluster and the 8 located within a few arcseconds of another SFiNCs source correspond to sources that also were not identified by D15, and are discussed in \S \ref{app:IR}, apart from serps1, serps3, and serps35 that have SFiNCs detections but not D15 classifications. 

We note that the two isolated, weak continuum detections (serps31 and serps4) with no SFiNCs counterpart are the same two sources mentioned in \S \ref{app:IR} that also did not have IR counterparts identified by D15.  As previously stated (\S \ref{app:IR}), these are more likely than the others to be non-YSO contaminants.  

We do not further delve into the agreement (or not) of the IR detections and classifications presented by \citet{Get17} and \citetalias{Dun15}.  We simply make the generalization that apart from the most clustered continuum sources, the SFiNCs classification method seems to confirm the young status of YSO members, and the continuum emission detected with ALMA likely corresponds to disks (according to the SFiNCs classifications).  Our continuum data show the power to resolve several sources into fragmented (binary or multiple) systems, and especially distinguish protostars in the very clustered central region, but classifying these deeply embedded sources may still be difficult with IR indicators.

\subsection{Sub-mm Detections with Herschel}
\begin{figure}[!ht] 
\includegraphics[width=\columnwidth,angle=0]{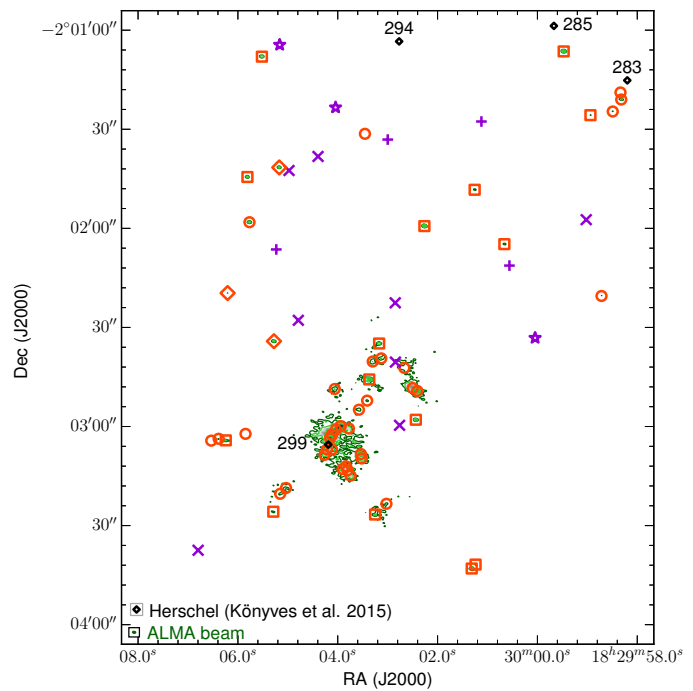}
\caption{ Locations of Herschel identified objects by \citet{Kon15} shown in black, with core running numbers from their Table A.2. Continuum map and non-black markers as in Figure \ref{fig:map}. }
\label{fig:herschel}
\end{figure}

We inspected the Hershel column density map and source identifications by \citet{Kon15}. While extended emission is prevalent, only 4 sources were identified by \citet{Kon15} within the coincident ALMA field of view (Figure \ref{fig:herschel}).  One of these sources (HGBS-J183004.1-020305, core 299) is near the central cluster and likely corresponds to the brightest continuum source serps45; it is classified as protostellar.  The extent of the emission in the column density map associated with this source encompasses the majority of compact sources we detect in the central region of the ALMA map (Figure \ref{fig:map_zoom}c). The other three Herschel sources are located in the northwest of the map near the edge, where our response is lower due to primary beam and mosaicking.  Two of these (HGBS-J182959.6-020058, core 285; and HGBS-J182958.2-020115, core 283) correspond within $\leq$10\arcsec to two continuum sources, serps7 and serps1, respectively.  These offsets are within the HPBW of the Herschel maps.  These sources are classified as protostellar and pre-stellar, respectively, by  \citet{Kon15}.

Non-detection in the Herschel images of most of the ALMA millimeter sources is likely limited by the angular resolution of Herschel.  Since most of the sources in the ALMA observations appear faint and compact, the signal may be diluted by the large beam size of the single-dish observations.  

\subsection{Radio sources} \label{sec:radio}

\citet{Ker16} made observations with the Karl G. Jansky Very Large Array (VLA) in two bands centered at 4.14 cm and 6.31 cm (beam sizes of $3.1\as \times 2.5$\arcsec and $4.8\as \times 3.8$\arcsec. respectively).  Radio emission probes high column densities, and can provide evidence of a central source that is young and embedded \citep{And00}.  The radio spectral index is powerful because it can in principle distinguish between thermal free-free emission and non-thermal radio emission (e.g., active galactic nuclei).

Twelve radio sources (VLA 6-17) coincide with our mapped region (see Table \ref{tab:sources} right column).  VLA 7, 11, 12, 13, 14, 16, 17 coincide precisely with 1.3 mm continuum detections (respectively: serps61, serps29, serps45, serps33, serps38, serps27, serps14).  VLA 7, 11, 16, 17 are classified by \citet{Ker16} as II, I, I, and F/II, respectively; these are all classified as Class 0/I by \citetalias{Dun15}, and while the discrepancy in spectral indices is indeed not clear to us, we can generally assume these are young YSOs (and moreover, not extragalactic).  VLA 12, 13, 14 coincide with mm sources in the dense central cluster, where no IR classifications were made.  VLA 12 likely coincides with the brightest 1.3 mm continuum source serps45, which also drives a collimated bipolar outflow \citep{Plu15b}, although within $4\as$\ are also several other mm sources along the mm continuum ridge (Figure \ref{fig:map_zoom}c).  It is heavily embedded in IR maps, and it is likely Class 0.  

\citet{Ker16} noted that VLA 13 is associated with an infrared source in all IRAC bands except the IRAC 1 band. They conclude based on fluxes at wavelengths spanning from IRAC 2 band (4.5 micron) to PACS 70$\mu$m as well as VLA detections that this source is class 0/I, and our ALMA detection (serps33) provides further evidence.  VLA 14 is also deeply embedded at IR wavelengths, and \citet{Ker16} note that it lies in projection within a gaseous envelope, according to observations from 24 to 250 $\mu$m.  It is likely an embedded Class 0 object.  

VLA 9 and VLA 10 are described by \citet{Ker16} as ``unclear'' sources.  These lie within $\lesssim4\as$\ of mm sources serps22 and serps26, respectively.  VLA 10 is elongated in the NW-SE direction, and hence might be one or two sources, indeed with the SE part most likely coincident with serps26, classified as class 0/I by D15.  Hence, we suggest that a protostar is located at the location of the southeastern part of VLA 10, while the northern part of VLA 10 (where we do not detect continuum emission) is either extragalactic or post-shock free-free emission from a neighboring outflow, possibilities introduced by \citet{Ker16}.

VLA 9 is near ($\sim4\arcsec$) to a weak IR source classified as class 0/I, but we do not detect emission with ALMA.  Its radio emission may blend with VLA 10, whose northern emission extends to within a few arcseconds of VLA 9, although VLA 9 appears less extended and more round than VLA 10.  Similar to VLA 10, it could be extragalactic, or it could be a second nearby instance of post-shock free-free emission, and it seems reasonable that two cases of this could appear in close proximity given the extent of outflow activity in that region \citep{Plu15a}.

VLA 6 and VLA 8 are significantly isolated and do not correspond with any ALMA continuum detection. VLA 6 is described as ``likely extragalactic''. VLA 8 is not discussed further by \citet{Ker16}, but listed as ``extragalactic ?''.  VLA 15 is located near to the central cluster in projection, but \citet{Ker16} classified it as extragalactic.

Interestingly, all of the ALMA continuum sources that are coincident with previous radio detections are classified as YSOs, and not extragalactic.  Hence by studying the radio map to supplement the ALMA map, we cannot determine that any of our mm source detections are in fact extragalactic contaminants.

\begin{figure}[!ht]
\includegraphics[width=\columnwidth,angle=0]{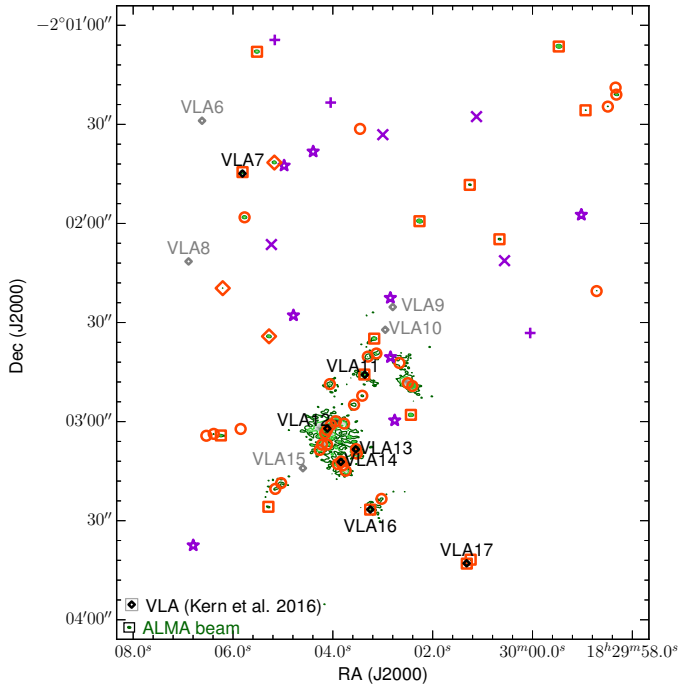} 
\caption{   Locations of VLA identified objects by \citet{Ker16} shown in black or gray, with respective labels from their table.  Black labels indicate sources classified as YSOs in \citet{Ker16}, while gray labels indicate a classification of ``unclear'' or ``extragalactic''.  See \ref{sec:radio} for full discussion. Continuum map and red/purple markers as in Figure \ref{fig:map}.}
\label{fig:vla}
\end{figure}

\end{document}